\documentclass[10pt]{article}   

\usepackage{amsthm,amsmath,amssymb,mathrsfs,cite,amscd}
\numberwithin{equation}{section}

\def\to{\mathchoice
{\longrightarrow}
{\rightarrow}
{\rightarrow}
{\rightarrow}}

\newcommand{\p}{\partial}

\newcommand{\half}{\tfrac12}

\newcommand{\CT}{\mathcal{T}}

\renewcommand{\geq}{\geqslant}

\newcommand{\C}{\mathbb{C}}

\newcommand{\Z}{\mathbb{Z}}

\renewcommand{\half}{\frac{1}{2}}
\newcommand{\tr}{{\rm tr}}

\renewcommand{\th}{\vartheta}
\newcommand{\bs}{\begin{split}}
\newcommand{\be}{\begin{equation}}
\newcommand{\es}{\end{split}} 
\newcommand{\ee}{\end{equation}}

\begin{document}


\newtheorem{defin}[equation]{Definition}
\newtheorem{lem}[equation]{Lemma}
\newtheorem{prop}[equation]{Proposition}
\newtheorem{thm}[equation]{Theorem}
\newtheorem{conj}[equation]{Conjecture}
\newtheorem{cor}[equation]{Corollary}
\newtheorem{rem}[equation]{Remark}
\newtheorem{ex}[equation]{Example}

\begin{titlepage} 
  \linespread{1.8}
  \title{\Large \bf N=4 Characters in Gepner Models, Orbits and
  Elliptic Genera.}
  \author{Daniel B. Gr\"unberg \\ {\em \small KdV Institute, Plantage
  Muidergracht 24, 1018 TV Amsterdam, The Netherlands} \\ \small grunberg@science.uva.nl} 
  \date{Apr 2003}
  \maketitle
  \vspace{2cm}
  \begin{abstract} \large
    We review the properties of characters of the N=4 SCA in the
    context of a non-linear sigma model on $K3$, how they are used to
    span the orbits, and how the orbits produce topological invariants
    like the elliptic genus.  We derive the same expression for the
    $K3$ elliptic genus using three different Gepner models ($1^6$,
    $2^4$ and $4^3$ theories), detailing the orbits and verifying that
    their coefficients $F_i$ are given by elementary modular
    functions.  We also reveal the orbits for the $1^3 2^2$, $1^4 4$
    and $1^2 4^2$ theories.  We derive relations for cubes of theta
    functions and study the function $ {1\over\eta}~ \sum_{n\in \Z}
    (-1)^n (6n+1)^k ~q^{(6n+1)^2 /24} $ for $k=1,2,3,4$.
  \end{abstract}
  \thispagestyle{empty}
\end{titlepage}
\thispagestyle{empty}  

\section{Introduction}
\label{sec:intro}

In the mid eighties, many efforts were deployed to find the characters
of the \underline{N=2} superconformal algebra (SCA), culminating in
the work of \cite{RY-87, G-88}.  The N=2 characters are defined by tr
$q^{L_0-c/24} y^{J_0}$ with $L_0$ the Virasoro operator and $J_0$ the
$U(1)$ charge.\footnote{We use the common variables $q=e^{2\pi i
    \tau}$ and $y=e^{2\pi i z}$ where $z$ keeps track of the $U(1)$
  theta angle.} They fall into two classes: those for continuous
central charge $c>3$ and those for discrete $c<3$, namely $c=3k/(k+2)$
with $k$ being the level.

In the first class, the characters for massive representations are
proportional to $\th_3 (z) / \eta^3$ (in the NS sector), while
those for massless representations have an extra denominator of
$1+y^{\rm{sign}(m)} q^{|m|-1/2}$, where $m$ is a quantum number
labelling the conformal dimension $h$ and the $U(1)$ charge $Q$.
Unitarity constrains $(h,Q)$ to lie inside a polygonal domain of the
plane.  Massless representations are those hitting the unitary bound,
ie with $(h,Q)$ on the boundary of the polygon.  Massive
representations are those with $(h,Q)$ in the interior of the polygon;
they have vanishing Witten index.

In the second class, with discrete $c<3$, the characters are spanned
by theta functions and the coefficients are the mysterious {\em string
 functions} of \cite{KP-84}.

In the late eighties, characters for the \underline{N=4} SCA were also
unravelled \cite{ET-88-1}.  The N=4 algebra contains an affine $su(2)$
Kac-Moody subalgebra of level $k$, and the central charge of the SCA
is $c=6k$.  For $k=1$, the massive characters are proportional to
$\th_3 (z)^2 / \eta^3$ (in the NS sector), while the massless
characters have again an extra denominator.  For higher level $k$, the
characters have an additional factor of $\chi^l_{k-1}$, which denotes
the $su(2)$ affine characters (or a slight deformation of them in the
massless case) with $l$ being the isospin quantum number, $0\leq l
\leq k/2$.  Unitarity requires $h\geq l$ (NS sector) and massless
representations hit this bound.  Massless N=4 characters were found to
be expressible as $\sum_{l'} A_{l,l'} \chi^{l'}_k$ for some branching
functions $A_{l,l'}$, and even expressible as an infinite sum of N=2
characters taken at double or triple points (these are special points
in the $(h,Q)$ plane).

This correspondence between N=2 and N=4 characters was furthermore
enhanced in \cite{EOTY-89}, where Gepner models were used to write N=4
characters -- or rather {\em orbits} -- as tensor products of several
characters of the N=2 minimal theories.  
Then finite sums over these orbits NS$_i$ or R$_i$ yield the traces
for the N=4 characters, modular invariant partition functions,
elliptic genera, or other topological invariants like
$$
\Phi= {\rm tr}^{^{\rm N=4}}_{_{\rm NS,R}} ~(-1)^F q^{L_0-c/24} y^{J_0} 
= \sum_i D_i ~{\rm NS}_i'~ \bar {\rm R}_i'
$$ 
for some combinatorial factors $D_i$.  Since these objects are
topological, they should not depend on the particular Gepner model at
hand.  Gepner models are special points in the moduli space of $K3$
surfaces \cite{NW-99} where the above trace factorizes into a product
of NS and R orbits.  That is, we only go to points where the formula
holds.  At different such points, we have different sets of orbits
NS$_i$ and of coefficients $D_i$.
Moreover, each orbit should be expressible as a sum of a
massless and a massive N=4 character: NS$_i(\tau,z) = \hat{\rm
  ch}^{\rm NS}(\tau,z) + F_i(\tau) ~{\rm ch}^{\rm NS}(\tau,z)$.  

In the context of $K3$ compactifications, the non-linear sigma model
has central charge $c=6$, thus the $su(2)$ subalgebra of the N=4 SCA
has level $k=1$.  In the following, we shall give explicit expression
of the functions $F_i(\tau)$ in the case of the $1^6$ and $2^4$
theories (and lay the cornerstone for the $4^3$ theory) and find that
they are essentially given by quotients of Dedekind $\eta$ functions,
thus reflecting the modular nature of the characters and topological
invariants.  We also derive the expression for $\Phi$ in both theories
and gather on the way useful results on theta functions and other
tools of analytic number theory.

This introduction is followed by six more sections.  In section 2, we
recall the N=4 characters for the $c=6$ SCA of the non-linear sigma
model with $K3$ target space. We also show how massless and massive
characters are used to span the {\em orbits}, without yet detailing
the construction of these orbits.  Section 3 is an expanded version of
the results of \cite{EOTY-89} on topological invariants for $K3$ based
on computations with the orbits.  Sections 4,5 and 6 are the crux of
the paper, revealing in detail the orbits for the $1^6$,$2^4$ and
$4^3$ Gepner models respectively, computing the functions $F_i$ and
developing several lemmas on theta functions.  Section 7 studies 
Gepner models of mixed levels, like $1^3 2^2$, $1^4 4$ and $1^2
4^2$ -- the first of which is a toroidal model and the other two are
$K3$ models.

In section 4, we also explore the function $a(\tau)$ which is
essential in \cite{BBG-94} for deriving Ramanujan identities.  In
particular, we study the function
$$
{1\over\eta}~ \sum_{n\in \Z} (-1)^n (6n+1)^k ~q^{(6n+1)^2 /24}
$$ 
for $k=1,2,3,4$ (Prop. \ref{a^2} and thereafter), and relate sums of
cubes of theta functions to a single theta function (lemma
\ref{cubic}).

\paragraph{Acknowledgements:} 
Many thanks to Anne Taormina for providing me with the orbits in
section \ref{sec:2^4-char-orb} and for further assistance, to Sander
Zwegers for the alternative proof of lemma \ref{cubic}, to Herschel
Farkas and Frank Garvan for their useful hints, to Katrin Wendland for
her feedback and helpful answers, and to Robbert Dijkgraaf for the
final touch.

\section{N=4 Characters}

We first write down the characters of the N=4 SCA with central charge
$c=6$ and level $k=1$, ie corresponding to a sigma model with $K3$
target space.  We give here explicitly the characters of the NS
sector, and refer to spectral flow for their counterparts in the R
sector.  They depend on two variables, $q=e^{2\pi i\tau}$ and
$y=e^{2\pi iz}$ for the modular parameter and the $U(1)$ theta angle
respectively.  Representations are parametrised by highest weight $h$
and isospin $l$ and unitarity implies $h \geq l$ (NS sector).  Our N=4
SCA is the enhancement of a N=2 Gepner model by adding $SU(2)$
currents $J^{\pm}$, and the latter's characters are defined by
\begin{equation}
  {\rm ch}^{\rm NS}(\tau,z) := \tr_{_{\rm NS}} ~q^{L_0-c/24} y^{J_0}
\end{equation}

\subsection{The Characters}

We rewrite the familiar expressions of \cite{ET-88-1} for N=4
characters in a more useful parametrisation.  There are two kinds of
characters: we denote massless characters (with isospin $l$) by
$\hat{\rm ch}^{\rm NS}_l (\tau,z)$ and massive ones (with highest
weight $h$) by ch$^{\rm NS}_h (\tau,z)$.

\underline{\bf Massless} representations saturate the unitarity bound:
$h=l$ and $l=0,\half$:
\begin{equation} \label{NS-char}
  \begin{split}
    \hat{\rm ch}^{\rm NS}_0 &= -{\th_3(z)\over \eta^3} \sum_n
    q^{n^2/2-1/8} y^n {1-yq^{n-1/2} \over 1+yq^{n-1/2}} 
    = 2\left( \th_1(z)\over \th_3 \right)^2 + \left(
    {q^{-1/8}\over\eta} -2 h_3 \right) \left(
    \th_3(z)\over \eta \right)^2 \\
     \hat{\rm ch}^{\rm NS}_\half &= {\th_3(z)\over \eta^3} \sum_n
    q^{n^2/2-1/8} y^n {1 \over 1+yq^{n-1/2}} 
    = -\left( \th_1(z)\over \th_3 \right)^2 + h_3~\left(
    \th_3(z)\over \eta \right)^2 ,
  \end{split}
\end{equation}
with 
\begin{equation}
  h_3 (\tau) := {1\over \eta \th_3}  \sum {q^{n^2/2-1/8} \over 1+q^{n-1/2}}
\end{equation}
The above equalities follow from the fact that the left hand sides are
so-called theta functions of characteristic $(0,0;-4\pi i, -2\pi
i\tau)$ of degree 2 (see appendix \ref{sec:theta-functions} below), hence
can be spanned by $\th_1(z)^2$ and $\th_3(z)^2$.  The coefficients are
obtained by evaluating the lhs at $z={1+\tau\over 2}$ and $z=0$ resp.,
bearing in mind that $\th_3({1+\tau\over 2})=0 $ and $\th_1(0)=0$.  For
$z={1+\tau\over 2}$, note that the term $(1-q^0)$ in the product expression
of $\th_3({1+\tau\over 2})$ cancels the denominator of the $n=0$ term of the
sum, yielding $2q^{-1/4}$ and $-q^{-1/4}$ for the left hand sides.

\underline{\bf Massive} representations are simpler and exist for $h>0$ and 
$l=0$:
\begin{equation} \label{massive-char}
  {\rm ch}^{\rm NS}_h = q^{h-1/8} ~{\th_3(z)^2\over \eta^3}.
\end{equation}

{\bf Spectral flow} yields the R character (idem for massive characters):
\begin{equation} \label{spectral-flow}
  \begin{array}{l}
\hat{\rm ch}^{\rm R}_l(\tau,z) = yq^{1/4} ~\hat{\rm ch}^{\rm
  NS}_{\half-l}~(\tau,z+{\tau\over 2}).
  \end{array}
\end{equation}
Thus for instance, the {\bf Witten index} is given by 
\begin{equation}
  \begin{array}{l}
  I= \tr_{_{\rm R}} ~q^{L_0-c/24} (-1)^F = \hat{\rm ch}^{\rm R}_{\half-l}
  (\tau,\half) = -q^{1/4} ~\hat{\rm ch}^{\rm NS}_l(\tau,{1+\tau\over 2}) = \left\{ 
  \begin{array}{ll}
-2, \quad &l=0\\
1, \quad &l=\half
  \end{array} \right.
  \end{array}
\end{equation}
since $\th_3({1+\tau\over 2}) =0$ and $\th_1({1+\tau\over 2})=q^{-1/8} \th_3$.
For the massive characters, the Witten index vanishes:
$\th_3({1+\tau\over 2}) =0$ in (\ref{massive-char}). 

\subsection{The Orbits}

The non-linear sigma models on $K3$ have three kinds of NS ``orbits'':
graviton, massless and massive orbits.  Their construction will be
detailed in the explicit computations below, sections
\ref{sec:1^6-theory}, \ref{sec:2^4-theory} and \ref{sec:4^3-theory}.
For now, we only need to know that they can be spanned by massless and
massive N=4 characters.  The graviton orbit, for example, contains the
massless character $ \hat{\rm ch}^{\rm NS}_0$ and a sum of massive
characters
$$
\sum_{n\geq 1} c_n ~{\rm ch}^{\rm NS}_n = \left( \sum_{n\geq 1} c_n
q^n \right) {\rm ch}^{\rm NS}_0 =: F_1(\tau) ~{\rm ch}^{\rm NS}_0.
$$ 
Thus the graviton orbit has coordinates $(1,F_1)$ in the basis $\{
\hat{\rm ch}^{\rm NS}_0, {\rm ch}^{\rm NS}_0 \} $.  Similarly for the
other orbits.  From the examples of the next sections, it will appear
that the massless orbits have always coordinates $(1,F_i)$, and the
massive orbits have coordinates $(0,F_j)$ (hence the name!).  We use
the subscripts $1,i,j$ for the different orbits: $1$ for the graviton
orbit, $i=2,\dots,d$ for the massless orbits and $j=d+1,\dots,d+d'$
for the massive orbits.  Writing the N=4 NS$_i$ characters (for the
three kinds of orbits) in the basis $\{ \hat{\rm ch}^{\rm NS}_{0,\half}, {\rm
  ch}^{\rm NS}_0 \} $ {\em defines} the functions $F_i (\tau)$ :
\begin{equation} \label{NS-orbits}
  \begin{array}{lll}
   {\rm NS}_1(\tau,z) &= \hat{\rm ch}^{\rm NS}_0(\tau,z) ~+& F_1(\tau) ~{\rm ch}^{\rm NS}_0(\tau,z)\\
   {\rm NS}_i(\tau,z) &= \hat{\rm ch}^{\rm NS}_\half(\tau,z) ~+& F_i(\tau) ~{\rm ch}^{\rm NS}_0(\tau,z)\\
   {\rm NS}_j(\tau,z) &=  &F_j(\tau) ~{\rm ch}^{\rm NS}_0(\tau,z)
  \end{array}
\end{equation}
The set of functions $F_i$ is determined by the particular Gepner model
under study.  Spectral flow generates again the Ramond counterparts,
R$_i$, and subsequent $(-1)^F $ insertion -- denoted by a prime --
yields the Witten index of the orbit (non-vanishing for massless
characters only): R$_1'= I_1=-2$, R$_i'= I_i=1$, R$_j'=0$.

The action of the modular group transforms all these orbits into each
other.  For instance, the $S$-transformation defines a real matrix
$S_{ij}$:
$$
{\rm NS}_i (\tau,z)= -\sum_j S_{ij}~{\rm NS}_j
\Big(-{1\over\tau},{z\over\tau}\Big) ~e^{-2\pi i~z^2/\tau}.
$$
Define $D_i:= S_{1,i}/S_{i,1}$, which are combinatorial factors of
tensoring representations when using a Gepner model.  For instance, in
the $1^6$ theory: $D_i=(1,20,270,30)$.  Using $D_i$, we form the
modular invariant partition function for the $K3$ $\sigma$-model: 
\begin{equation}
  Z(\tau,\bar \tau; z,\tau z)= {\rm tr}~ q^{L_0-{c\over 24}} \bar
    q^{\bar L_0-{\bar c\over 24}} y^{J_0} \bar y^{\bar J_0} =
    \half\sum_{i=1}^{d+d'} D_i \left( |{\rm NS}_i|^2 + |{\rm NS}'_i|^2
    +|{\rm R}_i|^2 +|{\rm R}'_i|^2 \right),
\end{equation}
where the prime represents $(-1)^F$ insertion.  The last term evaluated
at $y=1$ is but the Witten index $I$ and summing over it gives the
Euler character:
\begin{equation}
  \chi = \sum_{i=1}^{d+d'} D_i I_i^2= D_1 2^2 + D_2 + \dots D_d = 4+
  h^{1,1}= 24,
\end{equation}
as the sum of $D_i$ over the massless orbit always adds up to the Hodge
number $h^{1,1}$ of the orbifold: 20 in our case of $K3$.

\section{Topological invariants} \label{sec:topological-invariants}

\subsection{$K3$ elliptic genera}

In \cite{EOTY-89, KYY-93}, the authors studied the c=6 SCA of a
sigma-model with $K3$ target space.  The holonomy of the $K3$ manifold
allows for two more $SU(2)$ currents $J^\pm$, ie conformal fields of
weight 1 and $U(1)$ charge $J_0=2 J_0^3 =\pm 2 $, that generate the
transformation of double spectral flow (ie NS$ \to$R$\to$ NS) and
extend the N=2 algebra to N=4.

The elliptic genus of this $(c,\bar c)=(6,6)$ heterotic sigma model is,
geometrically, a double sum whose coefficients are the indices of
Dirac operators for certain vector bundles over $K3$:
$$
\begin{array}{rl}
\Phi (\tau,z) &= \sum_{n,r} c_{n,r} ~q^n y^r \\
 {\rm with} \quad c_{n,r} &:= {\rm ind}~ {\not\!\!D}_{E_{n,r}} =
 \int_{K3} {\rm ch}(E_{n,r}) ~{\rm td}(K3)
\end{array}
$$
where the bundle $E_{n,r}$ is defined by
$$ 
\sum_{n,r} E_{n,r}~ q^n y^r := y^{-1} \bigotimes_{n\geq 1}
\left( \bigwedge{ }_{-q^{n-1} y}T_{K3} \otimes \bigwedge{ }_{-q^{n-1}
    y^{-1}} \bar T_{K3} \otimes S_{q^n} T_{K3} \otimes S_{q^n}
  \bar T_{K3} \right),
$$ 
and  $ \bigwedge_{q} E = \bigoplus_{k\geq 0} q^k \bigwedge^k E $,  
$ S_{q} E = \bigoplus_{k\geq 0} q^k S^k E$.  ($\bigwedge^k$ and
$S^k$ denote the $k$'th exterior and symmetric products.)

The elliptic genus also has a topological expression, given by a trace
over the left and right Ramond sectors with $(-1)^F$ insertion:


\begin{equation}\label{ell-genus}
  \begin{split}
    \Phi (\tau,z) &:= \tr_{_{\rm R,R}} (-1)^{F_L+F_R} q^{L_0-1/4} \bar
  q^{\bar L_0-\bar 1/4} y^{J_0} \\
  &= 24\left(\frac{\th_3(z)}{\th_3}
  \right)^2 +2 \frac{\th_2^4-\th_4^4}{\eta^4} \left(
   {\th_1(z)\over\eta} \right)^2  ,
  \end{split}
\end{equation}
where the second expression will be proved in the next subsection.
Note that because we have no $\bar y^{\bar J_0}$ for the right
movers, the $(-1)^{F_R}$ insertion in the R-sector yields only a
contribution from the zero modes ($\bar q^0$-terms).  Indeed, for any
higher state, Susy ensures the existence of another state with
opposite $(-1)^F$ eigenvalue.  Thus the above expressions are
independent of $\bar q$ and we could have dropped that variable from
the definition.  Note also that the fermion parity operator $(-1)^{F}
=(-1)^{F_L+F_R} =(-1)^{F_L-F_R} $ is sometimes written $e^{i\pi
  (J_0-\bar J_0)}$.  The $U(1)$ charge $J_0$ helps to distinguish
between bosons and fermions, and its values are in $\Z$ for the NS
sector and in $\Z+\frac{c}{6}$ for the R sector.  Thus the difference
between left- and right-moving $U(1)$ charge is always an integer for
the NS-NS or R-R sectors.

At the special values of $z=\frac{1+\tau}{2}, \frac{\tau}{2},
\frac{1}{2}$ and $0$, we obtain specific topological invariants
\cite{EOTY-89}, using (\ref{t13}),(\ref{abstruse}) and dropping the
extra $q^{-1/4}$ and $-q^{-1/4}$ in the first two cases:
\begin{equation}  \label{top-invar}
  \begin{array}{llll}
{\rm Dirac\  index:}\qquad &\Phi^+_{\hat A} &:= \tr_{_{\rm NS,R}}
  (-1)^{F_R} q^{L_0-1/4} &= 2 \th_3^2
  (\th_2^4 -\th_4^4) / \eta^6 \\
&\Phi^-_{\hat A}&:= \tr_{_{\rm NS,R}} (-1)^{F_L+F_R} q^{L_0-1/4}  &= -2 \th_4^2
  (\th_2^4+\th_3^4)  / \eta^6 \\
{\rm Hirzebruch\ genus:}\qquad &\Phi_\sigma &:= \tr_{_{\rm R,R}} (-1)^{F_R}
  q^{L_0-1/4} &= 2 \th_2^2
  (\th_4^4+\th_3^4) / \eta^6  \\
{\rm Euler\  character:}\qquad &\Phi_\chi &:= \tr_{_{\rm R,R}}
  (-1)^{F_L+F_R} q^{L_0-1/4} &= 24
  \end{array}
\end{equation}
Whence a shift $z\to z+\frac{\tau}{2}$ generates spectral flow
R$\to$NS, while $z\to z+\frac{1}{2}$ is responsible for an additional
factor of $(-1)^{F_L}$.  The elliptic genus evaluated at specific
points thus yields the partition function for different spin
structures; at $z=0$, we obtain the Witten index -- or the bosonic
partition function if we have no spin structures.

We note that the above indices or genera are universal and do not
depend on the $K3$ moduli.  Since they hold for any complex structure,
they are rightly called topological invariants.  

\subsection{Derivation by Orbits}

We shall prove (\ref{ell-genus}) by actually computing $\Phi^+_{\hat
  A}$ with its $z$ dependence restored, ie we consider the NS,R
sector. This will allow us to work with the functions $F_i$ which we
defined by the left-moving NS$_i$ orbits.  Note that in the following,
the $\tau$-dependence shall be understood and not always explicitly
written. The prescription is to replace the trace by a sum over all
orbits:
$$
\Phi^+_{\hat A}(\tau,z) := \tr_{_{\rm NS,R}} q^{L_0-1/4} y^{J_0} (-1)^{F_R}
\bar q^{\bar L_0-\bar 1/4} = \sum_{i=1}^{d+d'} D_i ~{\rm NS}_i(\tau,z)
~\bar{\rm R}'_i(\bar\tau,\bar z =0)   
$$ 
This factorization of NS and R sectors will be confirmed by the
concrete examples of the next sections.\footnote{A thorough treatment
  of this factorization into tensor products of Hilbert spaces can be
  found in Katrin Wendland's PhD thesis \cite{W-00}}  Note that in
the right-moving sector, $ \bar{\rm R}'_i(\bar\tau,0)$ is but the Witten
index $I_1=-2$, $I_i=1$ and $I_j=0$.  Hence the trace consists of two
parts only, one for the graviton orbit and one from the massless
orbit.  In (\ref{NS-char}), we can interpret the coefficient of $
\th_1(z)^2/ \th_3^2$ as $-I_i$ for $i=1,\dots,d$ and similarly for the
coefficient of $h_3$.  Bearing this in mind, (\ref{NS-orbits}) gives
us:
\begin{equation*}
  \begin{split}
    \sum_{i=1}^d D_i  {\rm NS}_i(z) ~\bar{\rm R}'_i (0) &= \sum_i D_i \left[
    {\rm ch}^{\rm NS}_{0,\half}(z) + F_i ~{\rm ch}^{\rm NS}_0(z) \right] I_i \\
    &= \big(-\sum_{i=1}^d D_i I_i^2 \big) \left({\th_1(z)\over\th_3}
    \right)^2 + \bigg({q^{-1/8}\over\eta} \big(D_1 I_1 + \sum_{i=1}^d
    D_i I_i F_i \big) + h_3  
     \underbrace{\sum_{i=1}^d D_i I_i^2}_{=\chi=24} \bigg)  \left(
   {\th_3(z)\over\eta} \right)^2  
  \end{split}
\end{equation*}
Since this is a topological invariant, it should be independent of the
Gepner model at hand, ie of the particular set of functions $F_i(\tau)$.
That is, for different Gepner models we have different sets (of
variable length) of orbits NS$_i$ and functions $F_i$, but the above
sum yields always the same result.  In sections \ref{sec:1^6-theory} and
\ref{sec:2^4-theory} below, we show (for the $1^6$ and $2^4$ theories)
how the large bracket yields $2(\th_2^4-\th_4^4)/\eta^4$.  Hence our
Dirac index becomes:
\begin{equation} \label{dirac-index}
  \Phi^+_{\hat A}(z)=  -24 \left({\th_1(z)\over\th_3}
    \right)^2 + 2 {\th_2^4-\th_4^4 \over\eta^4} \left(
   {\th_3(z)\over\eta} \right)^2  
\end{equation}
and the $z=0$ value gives back the invariant of (\ref{top-invar}).

To arrive at the elliptic genus (\ref{ell-genus}), we need to insert $
(-1)^{F_L}$ and perform spectral flow for the left-movers.  This
corresponds to shifts $z\to z+\half$ and $z\to z+{\tau\over 2}$
respectively.  The first of these operations yields $\Phi^-_{\hat A}
(z) = \sum_{i=1}^d D_i {\rm NS}'_i(z) ~\bar{\rm R}'_i(0) $ and
combination with the second yields $\Phi(z) = \sum_{i=1}^d D_i
{\rm R}'_i(z)~\bar{\rm R}'_i(0)$ as in (\ref{ell-genus}).

\subsection{Alternative derivation by orbifolds} 

The expression for the elliptic genus (\ref{ell-genus}) can also
be derived from orbifold models of the $K3$ surface, as was shown in
\cite{EOTY-89}.  These models are formed by dividing the product of
two complex tori $T\times T'$ by the action of the symmetry group
$\Z_n$:
\begin{equation}  \label{orbifold-twists}
z_1 \to z_1 ~e^{2\pi i/n} \quad {\rm and } \quad z_2 \to
z_2 ~e^{-2\pi i/n}.  
\end{equation}
Essentially four types occur, corresponding to $n=2,3,4,6$.

The partition function for these models consists of an untwisted piece
and a twisted one.  The untwisted piece is the fermionic contribution
(in the NS sector, say) $|\th_3(z)/\eta|$ times the bosonic lattice
function $\Gamma_{2,2}(G,B)/|\eta|^4$.

The twisted piece consists of two complex fermions and two complex
bosons, twisted by some power of the $\Z_n$ symmetry generator
$e^{2\pi i/n}$, that is the $U(1)$ theta angle $z=2\pi \theta$ will be
shifted by $(s+r\tau)/n$.  For the fermions (in the NS sector, say),
we have again $\th_3(z)/\eta$ (yet with twists in opposite direction,
see (\ref{orbifold-twists})), while for the bosons we have
$\eta/\th_1$.  Thus the twisted partition function is the sum
\begin{equation}
  {\sum_{r,s}}' n_{r,s} |Z_{r,s}|^2, \qquad Z_{r,s}:=
  {\th_3(z+(s+r\tau)/n)~~\th_3(z-(s+r\tau)/n) \over
  \th_1((s+r\tau)/n)^2},
\end{equation}
where the prime on the sum signifies omission of $r=s=0$.  The weights
$n_{r,s}$ are defined by $n_{0,s}:= (s \sin(\pi s/n))^4/n$ and
$n_{r,s}:=n_{s,n-r}$.  Concretely, these weights all equal 8 for $n=2$
and 3 for $n=3$; while for $n=4$ the three weights forming at the
half-periods ($r,s=0,2$) equal 4 and the remaining twelve weights
equal 1.  For $n=6$, the three half-period weights equal 16/6, the
eight third-period weights equal 9/6 while the remaining twenty-four
weights equal 1/6.  In all cases, the important observation is that
the sum of the weights equals 24: $\sum_{r,s}' n_{r,s} =24$.

By the Riemann addition formula (\ref{riemann-addition}), the
$(r,s)$-block can be rewritten as
\begin{equation}
   Z_{r,s}(z)= \left( {\th_1(z)\over\th_3} \right)^2 + \left(
   {\th_3((s+r\tau)/n)\over \th_1((s+r\tau)/n)} \right)^2
   \left( {\th_3(z)\over \th_3} \right)^2 
\end{equation}
while its equivalent for the R-sector with $(-1)^F$ insertion is
\begin{equation}
   q^{1/4} y ~Z_{r,s}(z+(1+\tau)/2)=  \left(
   {\th_3(z)\over \th_3} \right)^2 - \left( 
   {\th_3((s+r\tau)/n)\over \th_1((s+r\tau)/n)} \right)^2
   \left( {\th_1(z)\over\th_3} \right)^2.
\end{equation}

With these building blocks, we can now compute the elliptic genus:
\begin{equation}
  \begin{split}
    \Phi(\tau,z) &= \tr_{_{\rm R,R}}  (-1)^F q^{L_0-1/4} y^{J_0}
    \bar q^{\bar L_0-\bar 1/4} \\
    &= {\sum_{r,s}}' n_{r,s} ~q^{1/4} y ~Z_{r,s}(z+(1+\tau)/2) ~~\bar
    q^{1/4}  ~\bar Z_{r,s}((1+\tau)/2) \\
    &= ({\sum_{r,s}}' n_{r,s})  \left( {\th_3(z)\over\th_3}
    \right)^2 - \left( {\sum_{r,s}}' n_{r,s} \left( 
   {\th_3((s+r\tau)/n)\over \th_1((s+r\tau)/n)} \right)^2 \right)
     \left( {\th_1(z)\over\th_3} \right)^2\\
     &= 24  \left( {\th_3(z)\over\th_3} \right)^2 
     + 2 {\th_2^4-\th_4^4 \over\eta^4} \left({\th_1(z)\over\eta}\right)^2 ,
  \end{split}
\end{equation}
where we have used (\ref{wp-theta}) to transform
\begin{equation}
  {\sum_{r,s}}' n_{r,s} \left( 
   {\th_3((s+r\tau)/n)\over \th_1((s+r\tau)/n)} \right)^2
     = \left({\th_3\over 2\pi \eta^3}\right)^2 {\sum_{r,s}}' n_{r,s}
   \bigg( \wp((s+r\tau)/n) -{\rm const } \bigg)
\end{equation}
and this last sum equals $-24 \cdot$const, by repeated use of
(\ref{wp-grid}) for equal values of $n_{r,s}$.  The constant itself
equals ${\pi^2\over 3}(\th_2^4-\th_4^4)$.  So we do indeed recover
(\ref{ell-genus}).

\section{Computations in $1^6$ Theory}
\label{sec:1^6-theory}

For clarity, we shall now detail the ideas developed at the beginning
of this section, and show what we mean under ``orbits'' and
functions $F_i$ in the concrete example of the $1^6$ theory.  This
Gepner model is based on the tensoring of six times the same $k=1$, N=2
SCFT.  That is, the N=4 characters will be tensor products of six N=2
characters.  So we first present those N=2 characters.

\subsection{General Considerations}

In general, for values of the central charge between 0 and 3, unitary
representations of N=2 superconformal algebras exist at discrete
values of the central charge, namely at $c=3k/(k+2)$.  The highest
weight states have conformal dimension and $U(1)$ charge parametrised
by two quantum numbers $l,m$ (isospin and its third component) \cite{RY-87}:
\begin{equation}
    h_{l,m}={l(l+2)-m^2 \over4 (k+2)}\qquad \qquad 
    Q_{l,m} ={m\over k+2}
\end{equation}
where $0\leq l \leq k,~~-l\leq m\leq l, ~~ l\equiv m$ mod
2. The NS characters of these N=2 theories are linear combinations of
$su(2)$ theta functions:
\begin{equation}
  \begin{split}
    {\rm ch}_{l,m}^{\rm NS}(y,q) &= \sum_{m'=-k+1}^k c_{l,m'}
  ~\theta_{(k+2)m'-mk, ~k(k+2)} \left( {\tau\over 2}, {z \over k+2}
  \right) \\
 \theta_{m,k} (\tau, z) &:= \sum_{n\in\Z +m/2k} q^{kn^2} y^{kn}, 
 \qquad \theta_{m,k} = \theta_{m+2k,~k}
  \end{split}
\end{equation}
For later purposes, note the behaviour under `full' ($z\to z+\tau$)
spectral flow: 
$$
\theta_{m, ~k(k+2)} \left( {\tau\over 2}, {z \over k+2} \right)
\stackrel{z\to z+\tau}{\longrightarrow} q^{-{k\over 2(k+2)}} y^{-{k\over (k+2)}}
~\theta_{m+2k, ~k(k+2)} \left( {\tau\over 2}, {z \over k+2} \right).
$$

The coefficients $c_{l,m}$ are the {\em string functions} of Kac and
Peterson \cite{KP-84} for $l\equiv m$ mod 2; for the $su(2)$ affine
Lie algebra they have an alternative definition via the  Weyl-Kac
formula:
\begin{equation}
  { \theta_{l+1,k+2}- \theta_{-l-1,k+2} \over \theta_{1,2}-
  \theta_{-1,2} } =: \sum_{m=-k+1}^k c_{l,m} \theta_{m,k}
\end{equation}
Since the lhs and rhs have expansions with powers of $y$ in $\Z+l/2$
and $\Z+m/2$ resp., we see that $c_{l,m}=0$ if $l\not\equiv m$
mod 2.  Of course, for each level $k$ we have different set of string
functions.  Note also the symmetries: $c_{l,m}=c_{l,-m} = c_{l,m+2k}
=c_{k-l,k-m}$.   For the case of the affine $su(2)$ algebra $A_1^{(1)}$,
  the string functions are merely proportional to Hecke `indefinite'
  modular forms:
\begin{equation}
    c_{l,m} = \eta(\tau)^{-3} \sum_{-|x|<y\leq |x|} {\rm sign}(x)
    ~q^{(k+2)x^2 -ky^2}
\end{equation}
where $x,y$ are such that $(x,y)$ or $(\half-x,\half+y)$ are  $\in
\Z^2 + ({l+1\over 2(k+2)},{m\over 2k})$. 

For our present case of $k=1$, $c=1$, the latter sum can be remarkably
rewritten as
\begin{equation}\label{eta-square-0}
      \begin{split}
   \sum_{-|x|<y\leq |x|\atop (x,y)\equiv ({1\over 6},0) {\rm
    or } ({1\over 3},\half) {\rm mod}\Z^2} {\rm sign}(x)~
    q^{3x^2 -y^2}  &= \sum_{j \geq 0\atop
      |l|\leq j/2} (-1)^{j+l} q^{(3(2j+1)^2-(6l+1)^2)/24}\\
  &= \eta(\tau)^2
      \end{split}
\end{equation}
The last equality is another remarkable result of \cite{KP-84}.  Our
string functions at level one thus become:
\begin{equation}
  c_{0,0}= c_{1,1} = c_{1,-1}= {1\over\eta(\tau)}, \qquad c_{0,1}= c_{1,0}= 0.
\end{equation}

\subsection{Characters and Orbits}

So we are in a position to write down the three minimal N=2
characters, obtained for $l=m=0$, $l=m=1$ and $l=-m=1$:
\begin{equation} \label{1^6-char}
  \begin{array}{lll}
A := {\rm ch}_{0,0}^{\rm NS}(y,q) = {1\over \eta}
\theta_{0,3}({\tau\over 2}, {z\over 3}) &= {1\over \eta}
\sum_\Z q^{{3\over 2} n^2} y^n &= {1\over \eta} ~\th_3(z|3\tau) \\
B := {\rm ch}_{1,1}^{\rm NS}(y,q) = {1\over \eta} 
\theta_{2,3}({\tau\over 2}, {z\over 3}) &= {1\over \eta}
\sum_\Z q^{{3\over 2} (n+{1\over 3})^2} y^{n+{1\over 3}} &= {1\over
  \eta} q^{1/6} y^{1/3}~ \th_3(z+\tau |3\tau)\\  
C := {\rm ch}_{1,-1}^{\rm NS}(y,q) = {1\over \eta} 
\theta_{4,3}({\tau\over 2}, {z\over 3}) &= {1\over \eta}
\sum_\Z q^{{3\over 2} (n+{2\over 3})^2} y^{n+{2\over 3}} &= {1\over
  \eta} q^{2/3} y^{2/3}~\th_3(z+2\tau |3\tau) \\
&&= {1\over \eta}q^{1/6} y^{-1/3}~ \th_3(z-\tau |3\tau)
  \end{array}
\end{equation}
Under spectral flow, the three $su(2)$ theta functions are shifted
into one each other:
\begin{equation}
  \begin{array}{l}
  \theta_{m,3}({\tau\over 2}, {z\over 3})
  \stackrel{z\to z+{\tau\over 2}}{\longrightarrow} q^{-1/24}
  y^{-1/6} ~\theta_{m+1,3}({\tau\over 2}, {z\over 3})
  \stackrel{z\to z+{\tau\over 2}}{\longrightarrow} q^{-1/6}
  y^{-1/3} ~\theta_{m+2,3}({\tau\over 2}, {z\over 3}),
  \end{array}
\end{equation}
so that under `full' spectral flow, the three characters are cyclicly
permuted: 
\begin{equation}
  A~~ \stackrel{z\to z+\tau}{\longrightarrow} ~~~B \to C \to A,
\end{equation}
where we have omitted the incrementing factors of $q^{-1/6}
  y^{-1/3} $.

To build the various orbits of the $1^6$ theory, we consider all
possible homogeneous polynomials of degree 6 in $A,B,C$, respecting
the following two rules: 
\begin{enumerate}
\item  the orbit must be holomorphic, ie its Fourier expansion must
  have integer powers of $y$;
\item  the orbit must be covariant under full spectral flow: NS$_i
  (z \to z+ \tau) = q^{-1} y^{-2}$ NS$_i (z)$.
\end{enumerate}
Condition (1) excludes combinations like $A^5 B$, $A^4 B^2$ or $A^3
B^2 C$,... Condition (2) requires invariance of the orbit under cyclic
permutation of $A,B,C$, and also guarantees it to be a theta function
of characteristic $(0,0;-4\pi i,2\pi i)$ and degree 2.  Thus each
orbit can be spanned by $\hat{\rm ch}_{0,\half}^{\rm NS}$ and
ch$_0^{\rm NS}$, or alternatively by $\th_1(z|\tau)^2$ and
$\th_3(z|\tau)^2$.

The four possible orbits respecting the above rules are:
\begin{equation} \label{1^6-orbits}
  \begin{split}
{\rm NS}_1 &= A^6 + B^6 +C^6 \\
{\rm NS}_2 &= A^3 B^3 + B^3 C^3 + C^3 A^3 \\
{\rm NS}_3 &= A^2 B^2 C^2 \\
{\rm NS}_4 &= A^4 B C + B^4 CA + C^4 A B .    
  \end{split}
\end{equation}

We recall the definition of the combinatorial factors $D_i$ associated
to a particular model:  The above orbits can be checked to have the
following modular behaviour \cite{EOTY-89}:
\begin{equation}\label{S_ij}
  {\rm NS}_i = -\sum_j S_{ij}~{\rm NS}_j
\Big(-{1\over\tau},{z\over\tau}\Big) ~e^{-2\pi i~z^2/\tau}, \qquad
S_{ij} = {1\over 27} \left( 
\begin{array}{rrrr}
3 & 60 & 270 & 90 \\
3 & -21 & 27 & 9 \\
1 & 2 & 9 & -6 \\
3 & 6 & -54 & 9
\end{array} 
 \right)
\end{equation}
Then the $D_i$ are defined by $D_i:= S_{1,i}/S_{i,1}$, that is
$(1,20,270,30)$ in the present case of $1^6$ theory.  Note that the
first column ($S_{i,1}$) is just the number of summands in the orbit
NS$_i$, while the first row ($S_{1,i}$) is $S_{1,1}$ times the number
of permutations of the factors in any summand of NS$_i$. The same
trick will allow a quick determination of the $D_i$ in the $2^4$ or
$4^3$ theories.

We note that at $y=-q^{-1/2}$, that is \underline{$z={1+\tau\over
    2}$}, the massive character vanishes and so does B, ch$(h=0)=0=B$,
    while $A=e^{i\pi /3} C =q^{-1/24}$.  Thus at $y=-q^{-1/2}$:
\begin{equation} \label{NS-special}
  \begin{array}{lll}
{\rm NS}_1 &= 2 q^{-1/4} &= \hat{\rm ch}_0^{\rm NS}(z={1+\tau\over 2}) \\
{\rm NS}_2 &= -  q^{-1/4}&= \hat{\rm ch}_{\half}^{\rm NS}(z={1+\tau\over 2}) \\
{\rm NS}_3 &= 0&\\
{\rm NS}_4 &= 0,&    
  \end{array}
\end{equation}
and we recognize that the first orbit is the graviton orbit, the
second is the massless orbit (only one), while the third and fourth
orbits are massive. 

\subsection{The functions $F_j$}

We will now compute the functions $F_j$ for the massive orbits.  For
$F_3$ this is pretty easy, while $F_4$ is more involved.  $F_1$ and
$F_2$ do not seem to have appealing expressions. 
Let us start with $F_3$:
\begin{equation}
  \begin{split}
{\rm NS}_3 &= A^2 B^2 C^2 = {q^{2/3} \over \eta^6} \big(
\th_3(z|3\tau) ~\th_3(z+\tau|3\tau) ~\th_3(z-\tau|3\tau) \big)^2 \\
&= {q^{2/3} \over \eta^6} \left( \th_3(z|\tau) ~\prod {(1-q^{3n})^3
  \over (1-q^n)} \right)^2 \\ 
&={\eta(3\tau)^6 \over \eta^8} ~\th_3(z|\tau)^2  \\
& \stackrel{!}{=} F_3~ {\rm ch}_0^{\rm NS} = F_3~q^{-1/8} ~{\th_3(z|\tau)^2
  \over \eta^3}
  \end{split}
\end{equation}
from which we find that 
\be
F_3 = q^{1/8} ~{\eta(3\tau)^6 \over \eta^5 }.
\ee
Similarly, for $F_4$ we have:
\begin{equation} 
  \begin{split}
{\rm NS}_4 &= ABC(A^3 + B^3 + C^3) \\
&=  {\eta(3\tau)^3
  \over \eta^4}  \th_3(z|\tau) ~\big[ \th_3(z|3\tau)^3 + q^{1/2} y~
\th_3(z+\tau|3\tau)^3  + q^{2/3} y^{2/3} \th_3(z+2\tau|3\tau)^3 \big]\\
& \stackrel{!}{=} F_4~ {\rm ch}(h=0) = F_4~q^{-1/8} {\th_3(z|\tau)^2
  \over \eta^3} .     
  \end{split}
\end{equation}
Thus we see that the large bracket with the sum of cubes of theta
functions must be proportional to $\th_3(z|\tau)$.  This is indeed the
content of lemma \ref{cubic} below, and we then obtain for $F_4$:
\be
F_4 = q^{1/8} ~{\eta(3\tau)^3 \over \eta^4} ~a(\tau),
\ee
where $a(\tau)$ is a function already studied in \cite{BBG-94}:
\begin{equation}
  \begin{split}
      a(\tau) &:={1\over\eta}~\sum_\Z (-1)^n (6n+1) ~q^{(6n+1)^2 /24}\\
      &= \sum_{k,l\in\Z} q^{k^2+kl+l^2}.
  \end{split}
\end{equation}

\subsection{Dirac Index}

Although we could not find interesting expressions for $F_1$ and
$F_2$, we shall nonetheless derive (\ref{dirac-index}), that is we
shall show:
\begin{prop}
  \begin{equation} \label{1^6-dirac-index}
  \sum_{i=1}^d D_i {\rm NS}_i I_i = -2 ~{\rm NS}_1 +20 ~{\rm
  NS}_2 = -24 ~{\th_1(z|\tau)^2 \over \th_3^2} + 2(\th_2^4-\th_4^4)
  ~{\th_3(z|\tau)^2 \over\eta^6}.
\end{equation}
\end{prop}
\begin{proof}
  Because the Dirac index (lhs) is spanned by the massless and massive
  characters ch$_0$ and ch, or equivalently by $\th_1(z|\tau)^2$ and
  $\th_3(z|\tau)^2$ , we only need to recover the constants
  multiplying these two basis vectors.  Due to (\ref{NS-special}) and
  the vanishing of $\th_3(z|\tau)$ at $z={1+\tau \over 2}$, we
  see that $\th_1(z|\tau)^2$ is correctly multiplied by
  $-24/\th_3^2$.  To check the constant in front of
  $\th_3(z|\tau)^2$ would only require setting $z=0$, where
  $\th_1(z|\tau)$ vanishes.  The lhs then would give $-2 (A^6+ 2~B^6)
  +20 (B^3 (2 A^3 +B^3)) $ because $B=C$ at $z=0$.  However, we have
  not succeeded in showing directly that this equals $
  2(\th_2^4-\th_4^4)~\th_3^2 /\eta^6$. Presumably, this is an
  interesting corollary of the theorem.

  Rather, to find the constants multiplying the two basis vectors, we
  shall differentiate both sides twice and set $z={1-\tau\over 2}$.
  This last evaluation has the merit of making the character $C$
  vanish, and giving also $A=-B={1\over\eta}~\th_3({1+\tau\over
    2}|3\tau)$. For NS$_1$, we have:
\begin{equation}
  \begin{split}
    \p_z^2\big|_{z={1-\tau\over 2}} ~{\rm NS}_1 
&= \p_z^2\big|_{z={1-\tau\over 2}} ~(A^6 + B^6 + C^6)  
= \p_z^2\big|_{z={1-\tau\over 2}} ~(A^6 + B^6) \\
&= 6 A^4 [ A''+5A'A] +6 B^4 [B'' +5 B'B].
  \end{split}
\end{equation}
Recalling that $\th_3$ and $\th_3''$ are even functions of $z$, while
$\th_3'$ is odd, and that they are all periodic under $z\to z+1$, we
note the following:  $\th_3({1-\tau\over 2}|3\tau) =
\th_3({1+\tau\over 2}|3\tau) $ and similarly for $\th_3''$,
but with an additional minus sign for  $\th_3'$.  Thus for instance, we
have at ${z={1-\tau\over 2}}$:
\begin{equation} 
    \begin{array}{ll}
      A &= \th_3(z|3\tau)\\ 
      A|_{z={1-\tau\over 2}} &= \th_3({1+\tau\over 2}|3\tau)\\ 
      A'|_{z={1-\tau\over 2}} &= -\th_3'({1+\tau\over 2}|3\tau)\\ 
      A''|_{z={1-\tau\over 2}} &=\th_3''({1+\tau\over 2}|3\tau) 
    \end{array} 
\end{equation} 
\begin{equation}
    \begin{array}{ll} 
      B &= q^{1/6} y^{1/3} ~\th_3(z+\tau|3\tau)\\ 
      B|_{z={1-\tau\over 2}} &= - \th_3({1+\tau\over 2}|3\tau)\\ 
      B'|_{z={1-\tau\over 2}} &= [-{2\pi i\over 3} \th_3
          -\th_3']({1+\tau\over 2}|3\tau) \\ 
      B''|_{z={1-\tau\over 2}} &= [ -{4\pi i\over 9} \th_3 +
         {4\pi i\over 3} \th_3' + \th_3'']({1+\tau\over 2}|3\tau)
     \end{array}
\end{equation}
In particular: $A'B+AB'= -{2\pi i\over 3} \th_3$. Bearing this in
mind, we obtain: 
\begin{equation}
  \begin{array}{l}
\p_z^2\big|_{z={1-\tau\over 2}} ~{\rm NS}_1 
= {1\over\eta^6} ~\th_3({1+\tau\over 2}|3\tau)^4 ~\big[
  12~\th_3''\th_3 + 60~{\th_3'}^2 + 48 \pi i~\th_3'\th_3 - 16\pi^2
  \th_3^2 \big] ({1+\tau\over 2}|3\tau). 
\end{array}
\end{equation}

Similarly, for NS$_2$, we have:
\begin{equation}
  \begin{array}{rl}
    \p_z^2\big|_{z={1-\tau\over 2}} ~{\rm NS}_2
&= \p_z^2\big|_{z={1-\tau\over 2}} ~(A^3 B^3 +B^3 C^3 +C^3 A^3 ) 
= \p_z^2\big|_{z={1-\tau\over 2}} ~(A^3 B^3 )\\ 
&= 3 A^2 B^2 ~[AB'' +2 A'B'+ A''B] + 6AB~[AB'+A'B]^2 \\
&=  {1\over\eta^6} ~\th_3({1+\tau\over 2}|3\tau)^4 ~\big[
-6 ~\th_3 \th_3'' + 6 ~{\th_3'}^2 +4\pi^2 ~\th_3^2 \big]
({1+\tau\over 2}|3\tau). 
  \end{array}
\end{equation}

Thus the {\bf lhs} altogether yields:
\begin{equation}
  \begin{array}{rl}
    \p_z^2\big|_{z={1-\tau\over 2}}( -2 ~{\rm NS}_1 +20 ~{\rm NS}_2)
&= -{4\over\eta^6} ~\big[  \th_3^5 ~(36 ~\th_3'' + {24\pi i}
~\th_3' -4\pi^2\th_3) -24\pi^2 ~\th_3 \big]({1+\tau\over 2}|3\tau) \\
&= 16 \pi^2 q^{-1/4} ~(6+E_2),
  \end{array}
\end{equation}
where we have noted that $\th_3({1+\tau\over 2}|3\tau)=q^{-1/24}
~\eta$ and that the curved bracket is proportional to the second
Eisenstein series $E_2 = {12\over i\pi} \p_\tau \log\eta$:
\begin{equation} \label{eisenstein}
  \begin{split}
    q^{1/24} (36 ~\th_3'' + {24\pi i} ~\th_3' -4\pi^2 ~\th_3)
&= -4\pi^2 \sum_\Z (-1)^n (6n+1)^2 ~q^{(6n+1)^2 /24} \\
&= -4\pi^2 {24\over2\pi i} ~\p_\tau \sum_\Z (-1)^n q^{(6n+1)^2 /24} \\
&= -4\pi^2 {24\over2\pi i} ~\p_\tau \eta \\
&= -4\pi^2 ~\eta ~E_2,
  \end{split} 
\end{equation}
by virtue of (\ref{t10}).

We now turn to the {\bf rhs} of (\ref{1^6-dirac-index}) and shall
differentiate twice.  To this effect, we remind a few useful facts:
\begin{equation}
  \begin{array}{llll}
    \th_1'(z+{1-\tau\over 2} | \tau) &= \p_z ~q^{-1/8} y^{1/2}
    ~\th_3(z) &= i\pi ~\th_1(z+{1-\tau\over 2}|\tau) + q^{-1/8} y^{1/2}
    ~\th_3'(z|\tau)   &\stackrel{z=0}{\longrightarrow} -i
    \pi q^{-1/8} ~\th_3 \\
    \th_3'(z+{1-\tau\over 2} | \tau) &= \p_z ~-i q^{-1/8} y^{1/2}
    ~\th_1(z) &= i\pi ~\th_3(z+{1-\tau\over 2}|\tau) -i q^{-1/8} y^{1/2}
    ~\th_1'(z|\tau) &\stackrel{z=0}{\longrightarrow} -
    2\pi i q^{-1/8} ~\eta^3, 
  \end{array}
\end{equation}
where we used (\ref{t11}).  Similarly, we find:
\begin{equation}
  \begin{array}{ll}
    \th_1''({1-\tau\over 2} | \tau) &= -\pi^2 q^{-1/8} ~\th_3 +
    q^{-1/8}  ~\th_3''(0|\tau) \\
    \th_3''(z+{1-\tau\over 2} | \tau) &= 4\pi^2 q^{-1/8} ~\eta^3 .
  \end{array}
\end{equation}
Thus equipped, we proceed for the rhs:
\begin{equation}
  \begin{split}
     \p_z^2\big|_{z={1-\tau\over 2}} \th_1(z|\tau)^2 
       &=  q^{-1/4} \th_3^2 ~[ -2 \pi^2 + \th_3'' / \th_3 ] \\
     \p_z^2\big|_{z={1-\tau\over 2}} \th_3(z|\tau)^2 
       &=  -4 \pi^2  q^{-1/4} ~\eta^6.
  \end{split}
\end{equation}
Taking (\ref{t16-biss}) into account, we have overall for the rhs:
\begin{equation}
 \p_z^2\big|_{z={1-\tau\over 2}}\left[ -24 {\th_1(z|\tau)^2
 \over \th_3^2} + 2(\th_2^4-\th_4^4) {\th_3(z|\tau)^2
 \over\eta^6} \right] 
=  16 \pi^2 q^{-1/4} ~(6+E_2),
\end{equation}
which was just the lhs.
\end{proof}

\subsection{Lemmas and Arithmetic Results}

\begin{lem}\label{cubic}
  \begin{equation}
    \begin{split}
\th_3(z|3\tau)^3 &+ q^{1/2} y ~\th_3(z+\tau |3\tau)^3 + q^{2} y^2
    ~\th_3(z+2\tau |3\tau)^3 = a(\tau) ~\th_3(z|\tau), \\
  {\rm where } ~a(\tau) &:={1\over\eta}~\sum_\Z (-1)^n (6n+1)
    ~q^{(6n+1)^2 /24} = \sum_{k,l\in\Z} q^{k^2+kl+l^2}.
     \end{split}
  \end{equation}
\end{lem}
\begin{proof}
That the rhs is proportional to $ \th_3(z|3\tau)$ follows from the
 second proof that we shall give.  To find the constant $a(\tau)$, 
we differentiate both sides wrt $z$ and set $z={1-\tau\over 2}$:
\begin{equation}
  \begin{array}{l}
   3 ~\th_3({1+\tau\over 2}|3\tau)^2 ~[-2~~\th_3' -{2\pi i \over
   3}~\th_3 ]({1+\tau\over 2}|3\tau) = a(\tau)~(-2\pi i
   q^{-1/8}~\eta^3)
  \end{array}
\end{equation}
Note that  $\th_3({1+\tau\over 2}|3\tau)=q^{-1/24} ~\eta$ and so we
find
\begin{equation}
  a(\tau)={1\over\eta}~\sum_\Z (-1)^n (6n+1) ~q^{(6n+1)^2/24}.
\end{equation}
For the second expression for $a$, we offer an alternative proof 
:
\begin{equation}
  {\rm lhs} = (\sum_\Z q^{{3\over 2}n^2} y^n)^3 +
  (\sum_{\Z+{1\over 3}} q^{{3\over 2}n^2} y^n)^3 +
  (\sum_{\Z+{2\over 3}} q^{{3\over 2}n^2} y^n)^3  
= \sum_n q^{{3\over 2} (n_1^2 + n_2^2 + n_3^2)} ~y^{(n_1 + n_2 + n_3)}
\end{equation}
where $n=(n_1,n_2,n_3)$ on the rhs sweeps through the set $S:=\Z^3 \cup
(\Z+{1\over 3})^3 \cup (\Z+{2\over 3})^3 $.  This set lies in 1
to 1 correspondence with all $k\in\Z^3$ via the following smart
substitution:
\begin{equation}
  \begin{array}{rl}
n_1&= (k_1 +k_2 -k_3)/3 \\
n_2&= (k_1 -2 k_2 -k_3)/3 \\
n_3&= (k_1 +k_2 +2k_3)/3 \\
  \end{array} \qquad
  \begin{array}{rl}
k_1&= n_1 +n_2 +n_3 \\
k_2&= n_1 -n_2  \\
k_3&= -n_1 +n_3 \\
  \end{array}
\end{equation}
Note that in the first definitions, all right hand sides are equal mod
1, which guarantees that all $n_i$ are in the same component of the
set $S$; whence the 1-1 correspondence. Moreover:
\begin{equation}
 n_1^2 + n_2^2 +n_3^2 = {1\over 3} k_1^2 + {2\over 3} (k_2^2
 +k_3^2 + k_2 k_3), \qquad \quad n_1 + n_2 + n_3 = k_1
\end{equation}
Hence
\begin{equation}
  {\rm lhs} =  \left( \sum_{k_2,k_3\in\Z} q^{k_2^2+k_2 k_3+k_3^2} \right)
  \sum_\Z q^{{1\over 2} k_1^2} y^{k_1}  =   {\rm rhs}
\end{equation}
\end{proof}

\begin{cor}
  \begin{equation}
  \begin{split}
 {\rm For} \quad z=0: &\quad \th_3(0|3\tau)^3 + 2q^{1/2} ~\th_3(\tau |3\tau)^3
 = a(\tau) ~\th_3 \\
 {\rm For} \quad z=1/2:& \quad \th_4(0|3\tau)^3 - 2q^{1/2} ~\th_3(\tau |3\tau)^3
 = a(\tau) ~\th_3 \\
 {\rm For} \quad z=3\tau/2: &\quad \th_2(0|3\tau)^3 + 2q^{1/4}
 ~\th_2(\tau|3\tau)^3 = a(\tau) ~\th_2 
  \end{split}
  \end{equation}
\end{cor}

For the sake of instruction, we give a third proof of the above
lemma, after reformulating it with a different constant of proportionality:
\begin{lem}\label{cubic-bis}
  \begin{equation}
    \begin{split}
\th_3(z|\tau)^3 &+ q^{1/6} y ~\th_3(z+\frac{\tau}{3} |\tau)^3
 + q^{2/3}  y^2 ~\th_3(z+{2\tau\over 3} |\tau)^3 = \Big( 6{\eta
    (3\tau)^3 \over\eta} +a(\tau)\Big)~\th_3(z|{\tau\over 3}), \\
  a(\tau) &:={1\over\eta}~\sum_\Z (-1)^n (6n+1) ~q^{(6n+1)^2 /24}
    \end{split}
  \end{equation}
\end{lem}
\begin{proof}
The advantage of having divided $\tau$ by 3 is that now all three terms on
lhs and the rhs are theta functions of degree 3 and characteristic
$(0,0;-6\pi i, 3\pi i\tau)$.  The space of such functions is 3
dimensional and can be spanned by 
$$
\begin{array}{l}
\Big\{ \th_3(z|{\tau\over 3}),~~ \th_3(z+{\tau\over 9}|{\tau\over 3}),~~
\th_3(z+{2\tau\over 9}|{\tau\over 3}) \Big\} ~~~{\rm or ~by }~~~~
\Big\{ \th_3(3z|3\tau), ~~y~\th_3(3z+\tau|3\tau), ~~y^2
~\th_3(3z+2\tau|3\tau) \Big\},
\end{array}
$$
etc.  Replacing $\tau\to 3\tau$, the lhs as a whole is still a theta
function of degree 1 and characteristic $(0,0;-2\pi i, \pi i\tau)$,
hence must be proportional to $\th_3(z|\tau)$. That is, all we have to
do is to compute the constant of proportionality.  To this end, we set
$z=0$ in the lemma and prove:
\begin{equation}
  \th_3(0|\tau)^3 +2 q^{1/6}~\th_3(\tau/3|\tau)^3 = \Big( 6{\eta
    (3\tau)^3 \over\eta} +a(\tau)\Big)~\th_3(0|{\tau\over 3})
\end{equation}

We quote from \cite{FK-01}, p. 273
, a property describing how the cubes of theta functions can be
spanned by basis vectors:
\begin{equation}
  \begin{array}{rl}
{i\pi \over 3} e^{{i\pi\over 6}} \eta~\th_3(z+{1+\tau\over
    2}|\tau)^3 =& -\pi e^{{i\pi\over 6}} q^{-5\over 24} y^{-{3\over
    2}} \eta(3\tau)^3~ \Big[ z^{1/2}~\th_3(3z+{1+\tau\over 2}|3\tau) -
    z^{-1/2}~\th_3(-3z+{1+\tau\over 2}|3\tau) \Big] \\
  &+ \th' ~\th_3(3z+3{1+\tau\over 2}|3\tau),
  \end{array}
\end{equation}
where 
\begin{equation}
  \begin{split}
      \th'&:= \th'[^{1/3}_{1}](0|3\tau):= \p_z|_0 ~~e^{2\pi i/3}
  q^{1/6} y^{-1/3} ~\th_1(z-\tau|3\tau), \\
{\rm such ~that }\quad -{3i \over\pi} e^{-i\pi/6}{\th'
  \over\eta} &= {1\over\eta}~\sum_\Z (-1)^n (6n+1) q^{(6n+1)^2
  /24} \\ &= a(\tau).
  \end{split}
\end{equation}
Special cases of this property are:
\begin{equation*}
  \begin{array}{rl}
    {\rm at } ~z=-{1+\tau\over 2}&: \quad \th_3(0|\tau)^3 = 6
    q^{1/6} {\eta (3\tau)^3 \over\eta}~ \th_3(\tau|3\tau)
    + a(\tau)~\th_3(0|3\tau) \\
    {\rm at } ~z=-\half +{\tau\over 6}&: \quad \th_3({\tau\over 3}|\tau)^3 = 3
   {\eta (3\tau)^3 \over\eta}~\Big[q^{-1/6}~\th_3(0|3\tau)
    +\th_3(\tau|3\tau) \Big]  + a(\tau)~\th_3(\tau|3\tau),
  \end{array}
\end{equation*}
so that 
\begin{equation}
    \th_3(0|\tau)^3 +2 q^{1/6}~\th_3(\tau/3|\tau)^3 = \Big( 6{\eta
    (3\tau)^3 \over\eta} +a(\tau)\Big)~ \Big[\th_3(0|3\tau) +
    2 q^{1/6}~\th_3(\tau|3\tau)\Big]
\end{equation}
Use lemma \ref{easy2} to rewrite the square bracket as $\th_3(0|\tau/3)$.
\end{proof}

Combining lemma \ref{cubic} and \ref{cubic-bis}, we arrive at an
interesting observation, already noticed in \cite{BBG-94}: 
\begin{cor}
  \begin{equation}
a(\tau/3) =  6{\eta (3\tau)^3 \over\eta} + a(\tau).    
  \end{equation}
\end{cor}

For completeness, we also observe that $a(\tau)$ can be written as
the difference of two Lambert series (ie $\sum {a_n q^n \over
  1-q^n}$): 
\begin{equation}
  \begin{split}
    a(\tau) &= {q^{1\over 24} \over\eta} \p_y|_1 ~\sum_\Z
    (-1)^n y^{6n+1} q^{n(3n+1)/2} \\
&= {q^{1\over 24} \over\eta} \p_y|_1 ~
    y~\th_3(6z+{1+\tau\over 2}|3\tau)  \\
&= {q^{1\over 24} \over\eta} \p_y|_1 ~ y~\prod
    (1-q^{3n})(1-y^{6n}q^{3n-1})(1-y^{-6n}q^{3n-2}) \\
&= 1+ 6 \sum_{n\geq 1} \left( {-q^{3n-1} \over 1-q^{3n-1}}+
    {q^{3n-2} \over 1-q^{3n-2}} \right)\\
&= 1+6\sum q^n \Big(\sum_{d|n \atop d\equiv 1 (3)} 1 - \sum_{d|n
    \atop d\equiv 2 (3)} 1 \Big) \\
&= 1+6\sum \delta_{3,1}(n)~q^n,
  \end{split}
\end{equation}
where $\delta_{k,l}(n)$ is the number of divisors of $n$ which are
${k-l \over 2}$ mod $k$ minus those which are ${k+l \over 2}$ mod $k$.
For example, a well-known result of Jacobi states that the number of
integer solutions to $x^2+y^2 = n$ is $4\delta_{4,2}(n)$.

Many more beautiful properties about $a(\tau)$ are found in
\cite{BBG-94}, such as 
\begin{equation}
  \begin{split}
  a(\tau)= \th_3(0|2\tau)~\th_3(0|6\tau)+
  \th_2(0|2\tau)~\th_2(0|6\tau).
\end{split}
\end{equation}

We give a last property, of our own, relating to
$a(\tau)^2$:
\begin{prop} \label{a^2}
\begin{equation} 
  2 a(\tau)^2  = 3 E_2(3\tau) -  E_2.
\end{equation}
\end{prop}
\begin{proof}
We apply the previous trick -- of differentiating a
Jacobi product -- to the sum already encountered in
(\ref{eisenstein}):
\begin{equation}
  \begin{split}
    E_2 &={1\over\eta}~ \sum_\Z (-1)^n (6n+1)^2 ~q^{(6n+1)^2 /24}\\
    &= {q^{1/24}\over\eta}~{-1\over 4 \pi^2}~\p_z^2|_0
    ~\sum(-1)^n y^{6n+1} q^{n(3n+1)/2}\\
    &= {q^{1/24}\over\eta}~{-1\over 4 \pi^2}~\p_z^2|_0
    ~y~\th_3(6z+{1+\tau\over 2}|3\tau) \\
 &= {q^{1/24}\over\eta}~{-1\over 4 \pi^2}~\p_z^2|_0
    ~y~\prod  (1-q^{3n})(1-y^{6n}q^{3n-1})(1-y^{-6n}q^{3n-2}) 
  \end{split}
\end{equation}
Abbreviating the last product by $\Pi$, we have that
$q^{1\over 24}\Pi|_0 =\eta$, ~$\Pi'=\Pi\Sigma$ and
$\Pi''=\Pi(\Sigma^2+\Sigma')$, where $\Sigma:= 12 \pi
i~\sum_{n\geq 1} \left(  {-y^6 q^{3n-1} \over 1-y^6 q^{3n-1}}+
    {y^{-6} q^{3n-2} \over 1-y^{-6} q^{3n-2}} \right)$.  In this
    notation, we also have $\Sigma|_0 =2\pi i (a(\tau)-1)$.  Thus:
\begin{equation}
  \begin{split}
    E_2 &= \Big[ 1+{1\over i\pi}~\Sigma -{1\over
    4\pi^2}(\Sigma^2+\Sigma') \Big]_0 \\
  &= 1+2(a-1) +(a-1)^2 -36~\sum_{n\geq 1} \left( {q^{3n-1} \over
    (1-q^{3n-1})^2} + {q^{3n-2} \over (1-q^{3n-2})^2} \right)\\
  &= a^2 +{3\over 2} \big[ E_2 -E_2(3\tau)\big].
  \end{split}
\end{equation}
\end{proof}

Since the sums ${1\over\eta}~ \sum_\Z (-1)^n (6n+1)^k ~q^{(6n+1)^2
  /24}$ yield enticing expressions for powers $k=1,2$ ($a(\tau)$,
$E_2$ resp.), it is natural to wonder whether this extends to higher
power.  We have not found any alternative expression for the case
$k=3$, but can nonetheless relate it to $a(\tau)$ and $E_2$.  Again,
we mimic the trick of the previous proposition:
\begin{equation}
  \begin{split}
    {1\over\eta}~ \sum_\Z (-1)^n (6n+1)^3 ~q^{(6n+1)^2 /24} 
    &= a^3 + {3a\over (2\pi i)^2} \Sigma'|_0 +{1\over (2\pi
      i)^3} \Sigma''|_0 \\ 
&=  a^3 + 3a~{3\over 2} (E_2 -E_2(3\tau))+{1\over (2\pi
      i)^3} \Sigma''|_0 ,
  \end{split}
\end{equation}
with 
\begin{equation}
\Sigma''|_0 =(2\pi i)^3 6^3 ~\sum_{n\geq 1} \left(
-{q^{3n-1}(1+q^{3n-1}) \over (1-q^{3n-1})^3} + {q^{3n-2}(1+q^{3n-2})
  \over (1-q^{3n-2})^3} \right) 
= (2\pi i)^3 6^3 ~\sum_{i\geq 1} i^2 ~{q^i(1-q^i) \over 1-q^{3i}}
\end{equation}

The same recursion for the case $k=4$ is even more involved and is not
worth writing in full, due to the complicated nature of
$\Sigma'''$:
\begin{equation}
  \begin{split}
    {1\over\eta}~ \sum_\Z (-1)^n (6n+1)^4 ~q^{(6n+1)^2 /24} 
    &= a^4 + {6a^2 \over (2\pi i)^2} \Sigma'|_0 +{4a\over (2\pi
      i)^3} \Sigma''|_0 + {3 \over (2\pi i)^4} {\Sigma'}|_0^2 + {1 \over (2\pi i)^4} \Sigma'''|_0\\ 
    &=  {1\over\eta}~ \left( {24\over 2\pi i} \p_\tau \right)^2
    \eta = 3 E_2^2 - 2 E_4.
  \end{split}
\end{equation}
The last line is obtained using the covariant derivative for modular
forms; it shows that $\p_\tau^{k/2} \eta$ (for $k$ even) can be
expressed as a polynomial in $E_2, E_4, E_6$.  For instance, for $k=6$
this is 
\begin{equation}
    {1\over\eta}~ \sum_\Z (-1)^n (6n+1)^6 ~q^{(6n+1)^2 /24} 
    = 16 E_6 - 30 E_2 E_4 + 15 E_2^3 .
\end{equation}

\section{Computations in $2^4$ Theory}
\label{sec:2^4-theory}

We mimic here the approach of the $1^6$ theory, as detailed in the
previous section.  This Gepner model is obtained by tensoring 4
times the $k=2$, N=2 theory.  Although we have more orbits and more cases to
study, the mathematics are easier, due to the simpler properties
enjoyed by theta functions with $\tau$ divided by 4 (instead of
divided by 3 as in the $1^6$ theory).

\subsection{Characters and Orbits:}\label{sec:2^4-char-orb}

This time we have six minimal N=2 characters, obtained for
$l=0~(m=0)$, $l=1 ~(m=\pm 1)$ and $l=2 ~(m=0,\pm 2)$.  The string
functions at level 2, due to their symmetries, number only three:
\begin{equation}
  c_{00}=c_{22}, \qquad c_{20}=c_{02} \qquad c_{1,-1}=c_{11}.
\end{equation}
In the following characters, we use the shorthand $\theta_m$ for the
$su(2)$ theta functions $\theta_{m,8} ({\tau\over 2}, {z\over 4})$:
\begin{equation} \label{2^4-char}
  \begin{array}{lll}
A := {\rm ch}_{0,0}^{\rm NS}(y,q) &= c_{00}~\theta_{0} +c_{02}~\theta_{8} 
   &= {1\over 2\eta} \sqrt{\th_3\over\eta} ~\th_3(z|2\tau)
   + {1\over 2\eta} \sqrt{\th_4\over\eta} ~\th_4(z|2\tau)\\ 
B := {\rm ch}_{2,2}^{\rm NS}(y,q) &= c_{02}~\theta_{-4} + c_{00}~\theta_{4}
   &= {1\over 2\eta} \sqrt{\th_3\over\eta} ~\th_2(z|2\tau) 
   + {1\over 2\eta} \sqrt{\th_4\over\eta} ~i ~\th_1(z|2\tau) \\ 
C := {\rm ch}_{2,0}^{\rm NS}(y,q) &= c_{02}~\theta_{0} +c_{00}~\theta_{8} 
   &= {1\over 2\eta} \sqrt{\th_3\over\eta} ~\th_3(z|2\tau)
   - {1\over 2\eta} \sqrt{\th_4\over\eta} ~\th_4(z|2\tau)\\ 
D := {\rm ch}_{2,-2}^{\rm NS}(y,q) &= c_{02}~\theta_{4}+c_{00}~\theta_{-4} 
   &= {1\over 2\eta} \sqrt{\th_3\over\eta} ~\th_2(z|2\tau) 
   - {1\over 2\eta} \sqrt{\th_4\over\eta} ~i ~\th_1(z|2\tau) \\ 
E := {\rm ch}_{1,-1}^{\rm NS}(y,q) &= c_{11}(~\theta_{-2} + \theta_{6})
   &= {\eta(2\tau)\over\eta^2} ~q^{1\over 16} y^{-{1\over
   4}} ~\th_3(z-{\tau\over 2} |2\tau)  \\ 
F := {\rm ch}_{1,1}^{\rm NS}(y,q) &= c_{11}(~\theta_{-6} +\theta_{2}) 
   &= {\eta(2\tau)\over\eta^2} ~q^{1\over 16} y^{1\over
   4} ~\th_3(z+{\tau\over 2} |2\tau) \\ 
  \end{array}
\end{equation}
The  $su(2)$ theta functions are related to the standard (or ``Ur-'')
Jacobi theta function via
\begin{equation}  
\begin{array}{l}
  \theta_{m,8} ({\tau\over 2}, {z\over 4}) = \sum_{n\in\Z}
  q^{(2n+{m\over 8})^2} ~y^{2n+{m\over 8}} = q^{({m\over 8})^2}
  ~y^{m\over 8} ~\th_3(2z+{m\tau\over 2}| 8\tau),\\
   \theta_m +  \theta_{m+8} = q^{({m\over 8})^2} y^{m\over 8}
  ~\th_3(z+{m\tau\over 4} |2\tau).
  \end{array}
\end{equation}
The rightmost column of (\ref{2^4-char}) is obtained by rewriting
$ac+bd$ as $\half (a+b)(c+d) +\half (a-b)(c-d)$ and using the 
explicit expression for the string functions from the next subsection.

Under spectral flow, these eight $su(2)$ theta functions are shifted
into one each other: 
\begin{equation}
  \begin{array}{l}
  \theta_{m,8}({\tau\over 2}, {z\over 4})
  \stackrel{z\to z+{\tau\over 2}}{\longrightarrow} q^{-1/6}
  y^{-1/4} ~\theta_{m+2,8}({\tau\over 2}, {z\over 4})
  \stackrel{z\to z+{\tau\over 2}}{\longrightarrow} q^{-1/4}
  y^{-1/2} ~\theta_{m+4,8}({\tau\over 2}, {z\over 4}),
  \end{array}
\end{equation}
so that under `full' spectral flow, the six characters split into two
groups which are cyclicly permuted: 
\begin{equation}
  A \stackrel{z\to z+\tau}{\longrightarrow} B \to C
  \to D \to A~,  \qquad \qquad E \to F \to E,
\end{equation}
where we have omitted the incrementing factors of $q^{-1/4}
y^{-1/2}$, etc.

To build the various orbits of the $2^4$ theory, we consider all
possible homogeneous polynomials of degree 4 in $A,B,C,D,E,F$, respecting
the following two rules: 
\begin{enumerate}
\item  the orbit must be holomorphic, ie its Fourier expansion must
  have integer powers of $y$;
\item  the orbit must be covariant under full spectral flow: NS$_i
  (z \to z+ \tau) = q^{-1} y^{-2}$ NS$_i (z)$.
\end{enumerate}
Note that $A,C$ have integer $y$-expansion, $B,D$ half-integer, and
$E,F$ have powers of $y$ in $\Z\mp {1\over 4}$ resp.  Thus,
condition (1) excludes combinations like $A^3 B,~A^3 E,~A^2 BC,~A^2
E$, etc.  Condition (2) requires invariance of the orbit under cyclic
permutation of $A,B,C,D$ and $E,F$ separately, and also guarantees it
to be a theta function of characteristic $(0,0;-4\pi i,2\pi i)$ and
degree 2.  Thus each orbit can be spanned by ch$_0(l=0,\half)$ and
ch$(h=0)$, or alternatively by $\th_1(z|\tau)^2$ and
$\th_3(z|\tau)^2$.

The twelve possible orbits respecting the above rules are:
\begin{equation*}
  \begin{array}{rllrl}
{\rm NS}_1 &= A^4 + B^4 +C^4 +D^4 & \qquad & {\rm NS}_7 &= AB^2C +BC^2D +CD^2A +DA^2B  \\
{\rm NS}_2 &= E^4 +F^4  & \qquad & {\rm NS}_8 &= ABCD  \\
{\rm NS}_3 &= A^2 B^2 + B^2 C^2 + C^2 D^2 + D^2 A^2 & \qquad &  {\rm NS}_9 &= ABE^2 +BCF^2 +CDE^2+DAF^2\\
{\rm NS}_4 &= ABF^2 +BCE^2 +CDF^2 +ADE^2  & \qquad & {\rm NS}_{10} &= (A^2 +B^2 +C^2 +D^2)EF  \\
{\rm NS}_5 &= B^2 D^2 + A^2 C^2  & \qquad & {\rm NS}_{11} &= E^2 F^2  \\
{\rm NS}_6 &= AC^3 +BD^3 +CA^3 +BD^3   & \qquad & {\rm NS}_{12} &= (AC + BD) EF \\
  \end{array}
\end{equation*}
The combinatorial factors $D_i$, defined after (\ref{S_ij}),
associated to these orbits are $(1, 2, 6, 12, 12, 4, 12, 96, 12,$ $12,
24, 48)$.  

Due to the following relations among the characters,
\begin{equation} \label{three-relations}
  \begin{array}{rl}
AC &= BD \\
AB + CD &= F^2 \\
AD + BC &= E^2,
  \end{array}
\end{equation}
(proved in lemma \ref{relations-char} with $\tau$ replaced by
${\tau\over 2}$), we find that several orbits coincide:
\begin{equation}
  \begin{array}{rl}
  {\rm NS}_2 &= {\rm NS}_4 \\ 
  {\rm NS}_5 &= 2~{\rm NS}_8 \\ 
  {\rm NS}_6 &=  {\rm NS}_7 = \half {\rm NS}_9 = {\rm NS}_{11}.
  \end{array}
\end{equation}

\subsection{String Functions:}

Some explicit expressions for the string functions at level 2 are found in
\cite{KP-84}, p. 220: 
\begin{equation}
  c_{11} = {\eta(2\tau)\over\eta^2}
  ={1\over\eta}{\sqrt{{\th_2\over 2\eta}}} , \qquad \qquad
  c_{00} -c_{02} = {\eta(\tau/2) \over\eta^2} =
  {1\over\eta}{\sqrt{{\th_4\over \eta}}} 
\end{equation}
The authors also give the complicated modular properties of the string
functions.  The latter expression, $c_{00} -c_{02}$, upon shifting
$\tau \to\tau +1$, yields $ e^{-i\pi/8} (c_{00} +c_{02}) $.
Similarly, shifting $ {1\over\eta}{\sqrt{{\th_4\over
      \eta}}}$ by $\tau \to\tau +1$ yields $ e^{-i\pi/8}
{1\over\eta}{\sqrt{{\th_3\over\eta}}} $.  Together with lemma
\ref{theta-8tau} below, this gives\footnote{The form
  ${1\over\eta}{\sqrt{{\th_i\over\eta}}}$ for the string functions
  $c_{11}$ and $c_{00} \pm c_{02}$ is expected from the fact that
  $\eta ~c_{ml}$ gives the character of the field $\phi_m^l$ in the $\Z_k$
  parafermion model \cite{FZ-85, GQ-87}.  For our case of $k=2$, the
  $\Z_2$ parafermion model is just the Ising model, and its characters
  are the well known square roots of theta functions (with different
  spin structures).  Thanks to Katrin Wendland for pointing at this and at
  her PhD thesis (p.50) \cite{W-00} which already contains the explicit
  expressions for (\ref{2^4-char}).}:
\begin{equation}
  \begin{split}
    c_{00} + c_{02} &= {1\over\eta}{\sqrt{{\th_3\over\eta}}} =
    {\eta\over\eta(2\tau) ~\eta(\tau/2)} 
    = {q^{-1/16}\over \th(\tau) -q^{1/2}~\th(3\tau) } \\
    c_{00} - c_{02} &= {1\over\eta}{\sqrt{{\th_4\over\eta}}} =
    {\eta(\tau/2) \over\eta^2}  
    = {q^{-1/16}\over \th(\tau) +q^{1/2}~\th(3\tau) }
  \end{split}
\end{equation}
Here, and for the remainder of this section, we use the shorthand
\underline{$\th(z):= \th_3(z|8\tau)$}.  Thus:
\begin{equation}
  \begin{split}
    c_{00} &= q^{-{1\over 16}} { \th(\tau) \over \th(\tau)^2 -q~\th(3\tau)^2}
        =q^{{1\over 16}} { \th(\tau) \over \eta ~\eta(2\tau)} \\
    c_{02} &= q^{-{1\over16}} { q^{1\over 2} ~\th(3\tau) \over
        \th(\tau)^2 -q~\th(3\tau)^2} 
        =q^{{1\over 16}} { q^{1\over 2} ~\th(3\tau) \over \eta ~\eta(2\tau)}
  \end{split}
\end{equation}
and
\begin{equation}
  \begin{split}
     c_{00}^2 - c_{02}^2 &= {1\over \eta ~\eta(2\tau)} ={ q^{-{1\over
     8}} \over \th(\tau)^2 -q~\th(3\tau)^2} \\
      c_{00}~ c_{02} &= q^{5\over 8} {\th(\tau)~\th(3\tau) \over
     \eta^2 ~\eta(2\tau)^2} = {\eta(8\tau)^2 \over \eta^3 ~\eta(4\tau)}
  \end{split}
\end{equation}

We now study the orbits at the special value of
\underline{$z={1+\tau\over 2}$}.  Note first that at this value,
$\theta_{m,8}({\tau\over 2}, {1+\tau\over 8}) = (-1)^{m\over 8}
q^{(({m\over 2})^2 +m)/16}~ \th_3(({m\over 2}+1)\tau |8\tau)$.  With
lemma \ref{theta-8tau}, our characters reduce to
\begin{equation}\label{NS-special-bis}
  \begin{array}{lll}
A = c_{00}~\th(\tau) +c_{02}~(-q^{3/2})~\th(5\tau) &=q^{-1/16}\\ 
C = c_{02}~\th(\tau) +c_{00}~(-q^{3/2})~\th(5\tau)&= 0\\ 
B = -i \big[c_{02}~\th(\tau) -c_{00} ~q^{1/2}~\th(3\tau) \big] &= 0\\ 
D = -i \big[c_{02}~\th(\tau) -c_{00}~q^{1/2}~\th(3\tau) \big] &=-i~q^{-1/16}\\ 
E = {\eta(2\tau)\over\eta^2} ~q^{-{1\over 16}} e^{-i\pi /4}
~\th_3(\half|2\tau) &=  e^{-i\pi /4} q^{-1/16} \\ 
F = {\eta(2\tau)\over\eta^2} ~q^{3\over 16} e^{-i\pi /4}
~\th_3(\half+\tau|2\tau) &= 0
  \end{array}
\end{equation}
Plugging these values into the orbits NS$_i$ yields NS$_1 = 2
q^{-1/4}$,  NS$_i = -q^{-1/4}$ ($i=2,\dots,4$), while the remaining
NS$_j$ vanish ($j=5,\dots,12$). We thus recognize from
(\ref{NS-orbits}) the graviton, massless and massive orbits
respectively.  

\subsection{The Functions $F_i$, $F_j$ and the Dirac Index}

In order to compute the functions $F_i$ (\ref{NS-char}), we set
\underline{$z=0$}, in which case $\theta_{m,8}({\tau\over 2},0) =
q^{({m\over 8})^2}~ \th({m\tau \over 2}) = \theta_{-m,8}$.  Hence the
characters simplify to:
\begin{equation}
  \begin{array}{rl}
A+C &= {1\over \eta} \sqrt{\th_3\over\eta} ~\th_3(0|2\tau) 
    = {\eta(2\tau)^4 \over\eta~\eta(4\tau)^2 ~\eta(\tau/2)}
    = E^2/B \\
A-C &= {1\over \eta} \sqrt{\th_3\over\eta} ~\th_4(0|2\tau) 
    = {\eta(\tau/2) \over\eta(2\tau)}= 1/E \\
B = D &=  {1\over \eta} \sqrt{\th_3\over\eta} ~\th_2(0|2\tau)  = {\eta
  ~\eta(4\tau)^2 \over \eta(\tau/2) ~\eta(2\tau)^2 } \\
E = F &= {\eta(2\tau) \over \eta(\tau/2)} = \sqrt{\half {\th_2\over\th_4}}
  \end{array}
\end{equation}
With these observation and the fact that $AC=B^2$ at $z=0$
(\ref{three-relations}), the \underline{massive} orbits at $z=0$ have
a rather simple form:
\begin{equation}
  \begin{array}{l}
{\rm NS}_5 = 2 {\rm NS}_8 = 2 B^4 \\
{\rm NS}_6 ={\rm NS}_7 = \half {\rm NS}_9 ={\rm NS}_{11} = E^4 \\
{\rm NS}_{10} = E^6/B^2 = {\rm NS}_{12} ~{\rm NS}_6 / {\rm NS}_5 \\
{\rm NS}_{12} = 2 B^2 E^2 
  \end{array}
\end{equation}
Given that NS$_j = F_j ~q^{-{1\over 8}}~{\th_3^2 \over
  \eta^3}=  F_j ~{q^{-{1\over 8}}~\eta^7 \over \eta(2\tau)^4~
  \eta(\tau/2)^4 }$ , we find the following values for $F_j$:
\begin{equation}
  \begin{array}{l}
F_5 = 2 F_8 =  2q^{1/8} {\eta(4\tau)^8 \over \eta^3 ~\eta(2\tau)^4 } \\
F_6 = F_{7} = \half F_9 =F_{11} = q^{1/8} {\eta(2\tau)^8 \over \eta^7} \\
F_{10} =  q^{1/8} {\eta(2\tau)^{14} \over \eta^9 ~\eta(4\tau)^4}\\
F_{12} = 2q^{1/8} { \eta(2\tau)^2 ~\eta(4\tau)^4 \over \eta^5 }\\
F_5  F_{10} = F_6 F_{12}
  \end{array}
\end{equation}

The \underline{massless} orbits are a little less elegant, especially
NS$_1$ and NS$_3$ which are not factorizable.  For the latter, we
shall need the following observation, again having set $z=0$:
\begin{equation}
  \begin{split}
(A^2 + C^2)  &= (A-C)^2 +2B^2 =1/E^2 + 2B^2 \\
       &= (A+C)^2 -2B^2 = E^4/B^2 - 2B^2 \\
\Rightarrow & E^6 =B^2 + 4 B^4 E^2
  \end{split}
\end{equation}
and 
\begin{equation}
   \begin{split}
   1/E^2 + 2B^2 &= 2 {\th_4 \over \th_2} + {\th_2(0|2\tau)^2
    \over \th_2 ~\th_4} = {2\th_4^2 +\th_2(0|2\tau)^2
    \over \th_2 ~\th_4} =  {3\th_4^2 + \th_3^2  \over 2~\th_2 ~\th_4}
  \end{split}
\end{equation}
We note that the last expression cannot be factorized
and so we give the graviton + massless orbits as they stand:
\begin{equation}
  \begin{array}{ll}
{\rm NS}_1 &= (A^2 +C^2)^2 = (1/E^2 + 2B^2)^2 = \left( {3\th_4^2
  +\th_3^2  \over 2~\th_2 ~\th_4} \right)^2 \\
{\rm NS}_2 = {\rm NS}_4 &= 2 ~E^4 \\
{\rm NS}_3 &= (3~\th_4^2 +\th_3^2) ~{\th_3^2 ~\th_2(0|2\tau)^2
  \over 8 \eta^6} = (3~\th_4^2 +\th_3^2) ~{\th_3^2 -\th_4^2
  \over 4~\th_2^2~\th_4^2} =  (3~\th_4^2 +\th_3^2)~ {\eta
  ~\eta(4\tau) \over  2~\eta(\tau/2) ~\eta(2\tau)}\\ 
{\rm NS}_3^2 &= 2 ~{\rm NS}_1 {\rm NS}_5
  \end{array}
\end{equation}
with extensive use of formulae in appendix \ref{app-theta}.  Given
that NS$_i = (h_3+F_i ~{q^{-{1\over 8}}\over \eta}) ~{\th_3^2 \over
  \eta^2}$ for $i=2,3,4$ (and $h_3$ replaced by $-2 h_3$ for NS$_1$) ,
we find the following values for $F_1$ and $F_i$:
\begin{equation}
  \begin{array}{l}
F_1 = q^{1/8} ~\eta \Big[ \left( {3~\th_4^2 +\th_3^2  \over
     4~\eta^2 }\right)^2  + 2~h_3 \Big] -1 \\
F_2 = q^{1/8} ~\eta \Big[ 2 \left( {\eta(2\tau) \over
     \eta } \right)^8 -h_3  \Big]  = 2 ~F_6 - q^{1/8} ~\eta ~h_3 \\
F_3 =  q^{1/8} ~\eta \Big[ (3~\th_4^2 +\th_3^2) {\eta(4\tau)^4 \over
     2~\eta^4 ~\eta(2\tau)^2 } -h_3 \Big] \\
F_4 = F_2
\end{array}
\end{equation}

As in the $1^6$ theory, we shall again derive (\ref{dirac-index}), that is we
shall show:
\begin{prop}
  \begin{equation} \label{2^4-dirac-index}
  \sum_{i=1}^d D_i {\rm NS}_i I_i = -2 ~{\rm NS}_1 +2 ~{\rm
  NS}_2 + 6~{\rm NS}_3 +12~ {\rm NS}_4 = -24 ~{\th_1(z|\tau)^2 \over
  \th_3^2} + 2(\th_2^4-\th_4^4)   ~{\th_3(z|\tau)^2 \over\eta^6}.
\end{equation}
\end{prop}

\begin{proof}
Due to (\ref{NS-special-bis}) and the vanishing of $\th_3(z|\tau)$
  at $z={1+\tau \over 2}$, we see that $\th_1(z|\tau)^2$ is
  correctly multiplied by $-24/\th_3^2$.
  We only need to recover the factor multiplying $\th_3(z|\tau)^2$.
  Unlike in $1^6$ theory, setting  $z=0$ on both sides will easily do
  the job:
\begin{equation}
    {\rm lhs}|_0 = {1\over 2~\th_2^2 ~\th_4^2} \Big[
-(3~\th_4^2+\th_2^2)^2 +(2+12)~ \th_2^4 +6(3~\th_4^2+\th_3^2)
\half (\th_3^2-\th_4^2) \Big] 
  \end{equation}
The square brackets yield a total of $16 (\th_2^4 - \th_4^4)$,
while the prefactor is $\half \left( {\th_3 \over 2~\eta^3}
\right)^2$ .  Thus we obtain the rhs.
\end{proof}

\subsection{Lemmas:}

\begin{lem}\label{theta-8tau}
With the shorthand $\th(z):= \th_3(z|8\tau)$, we have:
    \begin{enumerate}
\item $ q~\th(5\tau) = \th(3\tau)$
\item $\th(0) -q~\th(4\tau) = \th_4 (0|2\tau) = {\eta^2 \over \eta(2\tau)}$
\item $\th(0) +q~\th(4\tau) = \th_3 (0|2\tau) = {\eta^2 \over \eta(2\tau)}$
\item $\th(\tau) +q^{1/2}~\th(3\tau) = q^{-{1\over 16}} \half \th_2(0|\tau/2) = q^{-{1\over 16}} {\eta^2 \over \eta(\tau/2)} $
\item $\th(\tau) -q^{1/2}~\th(3\tau) = \th_2(\tau/2|2\tau) =
  q^{-{1\over 16}}  {\eta(2\tau) ~\eta(\tau/2) \over \eta } $
    \end{enumerate}
\end{lem}

\begin{proof} These are all instances of the more general lemma
  \ref{easy1}.  Alternatively:
  \begin{enumerate}
  \item directly from sum or product expression.
  \item lhs $= \sum q^{4n^2}- \sum q^{4(n+\half)^2} =
    \sum q^{4({n\over 2})^2} = \th_4 (0|2\tau) =
    \prod(1-q^{2n})(1-q^{2n-1})^2 =\prod(1-q^{n})(1-q^{2n})^{-1} =$rhs
  \item  idem
   \item lhs $= q^{-{1\over 16}} \sum \big( q^{4(n-{1\over
         8})^2} +q \sum q^{4(n+{3\over 8})^2} \big) =  q^{-{1\over 16}} 
        \sum q^{(n-{1\over 4})^2} =  q^{-{1\over 16}} \half
        \sum q^{(n-\half)^2} = q^{-{1\over 16}} \half \th_2(0|\tau/2) =
        \prod(1-q^{n/2})(1+q^{n/2})^2 = \prod(1-q^n)^2
        (1-q^{n/2})^{-1} =$ rhs $=  \th_3(\tau/2|2\tau) $
  \item  idem
  \end{enumerate}
\end{proof}

\begin{lem}\label{relations-char}
  \begin{equation}
      \begin{array}{l}
 \th_4(0|{\tau\over 2}) ~\big( \th_1(z|\tau)^2 + \th_4(z|\tau)^2 \big) 
= \th_3(0|{\tau\over 2}) ~\big( \th_3(z|\tau)^2 - \th_2(z|\tau)^2 \big) 
= \th_3(0|{\tau\over 2}) ~\th_3({z\over 2}|{\tau\over 4})
\th_4({z\over 2}|{\tau\over 4}) \\
  \th_3(0|{\tau\over 2}) ~\th_2(z|\tau) \th_3(z|\tau) 
+ \th_4(0|{\tau\over 2}) ~i \th_1(z|\tau) \th_4(z|\tau) 
= \th_2(0|{\tau\over 2}) ~q^{1\over 16} y^{-\half}~
\th_3(z-{\tau\over 4}|\tau)^2 \\
  \th_3(0|{\tau\over 2}) ~\th_2(z|\tau) \th_3(z|\tau) 
- \th_4(0|{\tau\over 2}) ~i \th_1(z|\tau) \th_4(z|\tau) 
= \th_2(0|{\tau\over 2}) ~q^{1\over 16} y^\half ~
\th_3(z+{\tau\over 4}|\tau)^2 
  \end{array}
  \end{equation}
\end{lem}

\begin{proof}
Both sides of the first equation are theta functions of degree two and
characteristic $(0,0; -4\pi i, -2\pi i\tau)$, ie elements of the
two-dimensional vector space $\CT_{2, -2\pi i\tau}$ (see appendix
\ref{sec:theta-functions}).  In the other two equations, all terms  --
upon extra multiplication with $y$ -- are elements of $\CT_{2, -3\pi
  i\tau}$.  So all we need to do is to verify the relations at two
independent values of $z$. 

For all line, verification at $z=0,\tau/2$ is immediate with
(\ref{dupl-1}), (\ref{dupl-2}). The rhs gives a handy factorized form
of the lhs, which would not lend itself to straightforward
factorisation as the sums $(i\th_1 \pm \th_4)$ cannot be made
into a theta function.  
\end{proof}


\section{Computations in $4^3$ Theory}
\label{sec:4^3-theory}

We mimic here the approach of the two previous sections.  This Gepner
model is obtained by tensoring 3 times the $k=4$, N=2 theory.  We have
considerably more orbits and more cases to study; the mathematics
involve properties of theta functions with $\tau$ divided by 3, 4 and 6.

\subsection{Characters and Orbits:}

This time we have 15 minimal N=2 characters, obtained for
$l=0~(m=0)$, $l=1 ~(m=\pm 1)$, $l=2 ~(m=0,\pm 2)$, $l=3 ~(m=\pm1,3)$, 
$l=4 ~(m=0,\pm 2, \pm 4)$.  The string
functions at level 4, due to their symmetries, number only seven:
\begin{equation}
  c_{00}=c_{44}, \qquad c_{02}=c_{42}, \qquad c_{04}=c_{40}, \qquad
  c_{20}=c_{24},  \qquad c_{22}, \qquad c_{11}=c_{33}, \qquad c_{13}=c_{31}.
\end{equation}

In the following characters, $\theta_m$ is a shortcut for the
$su(2)$ theta function evaluated as $\theta_{m,24} ({\tau\over 2},
{z\over 6})$:
\begin{equation}
  \begin{array}{lllll}
A := {\rm ch}_{0,0}^{\rm NS}(y,q) &= c_{02}~\theta_{-12} &+
c_{00}~\theta_{0} &+ c_{02}~\theta_{12} &+ c_{04}~\theta_{24}\\ 
B := {\rm ch}_{4,4}^{\rm NS}(y,q) &= c_{02}~\theta_{20} &+
c_{04}~\theta_{-16} &+ c_{02}~\theta_{-4} &+ c_{00}~\theta_{8}\\ 
C := {\rm ch}_{4,2}^{\rm NS}(y,q) &= c_{02}~\theta_{-20} &+
c_{04}~\theta_{-8} &+ c_{02}~\theta_{4} &+ c_{00}~\theta_{16}\\ 
D := {\rm ch}_{4,0}^{\rm NS}(y,q) &= c_{02}~\theta_{-12} &+
c_{04}~\theta_{0} &+ c_{02}~\theta_{12} &+ c_{00}~\theta_{24}\\ 
E := {\rm ch}_{4,-2}^{\rm NS}(y,q) &= c_{02}~\theta_{-4} &+
c_{04}~\theta_{8} &+ c_{02}~\theta_{20} &+ c_{00}~\theta_{-16} \\
F := {\rm ch}_{4,-4}^{\rm NS}(y,q) &= c_{02}~\theta_{4} &+
c_{04}~\theta_{16} &+ c_{02}~\theta_{-20} &+ c_{00}~\theta_{-8}\\
  \end{array}
\end{equation}
\begin{equation}
  \begin{array}{lllll}
G := {\rm ch}_{2,2}^{\rm NS}(y,q) &= c_{22}~\theta_{-20} &+
c_{20}~\theta_{-8} &+ c_{22}~\theta_{4} &+ c_{20}~\theta_{16}\\ 
H := {\rm ch}_{2,0}^{\rm NS}(y,q) &= c_{22}~\theta_{-12} &+
c_{20}~\theta_{0} &+ c_{22}~\theta_{12} &+ c_{20}~\theta_{24}\\ 
I := {\rm ch}_{2,-2}^{\rm NS}(y,q) &= c_{22}~\theta_{-4} &+
c_{20}~\theta_{8} &+ c_{22}~\theta_{20} &+ c_{20}~\theta_{-16} \\
  \end{array}
\end{equation}
\begin{equation}
  \begin{array}{lllll}
J := {\rm ch}_{1,1}^{\rm NS}(y,q) &= c_{13}~\theta_{-22} &+
c_{11}~\theta_{-10} &+ c_{11}~\theta_{2} &+ c_{13}~\theta_{14}\\ 
K := {\rm ch}_{1,-1}^{\rm NS}(y,q) &= c_{13}~\theta_{-14} &+
c_{11}~\theta_{-2} &+ c_{11}~\theta_{10} &+ c_{13}~\theta_{22}\\ 
L := {\rm ch}_{3,3}^{\rm NS}(y,q) &= c_{11}~\theta_{18} &+
c_{13}~\theta_{-18} &+ c_{13}~\theta_{-6} &+ c_{11}~\theta_{6}\\ 
M := {\rm ch}_{3,1}^{\rm NS}(y,q) &= c_{11}~\theta_{-22} &+
c_{13}~\theta_{-10} &+ c_{13}~\theta_{2} &+ c_{11}~\theta_{14}\\ 
N := {\rm ch}_{3,-1}^{\rm NS}(y,q) &= c_{11}~\theta_{-14} &+
c_{13}~\theta_{-2} &+ c_{13}~\theta_{10} &+ c_{11}~\theta_{22}\\ 
O := {\rm ch}_{3,-3}^{\rm NS}(y,q) &= c_{11}~\theta_{-6} &+
c_{13}~\theta_{6} &+ c_{13}~\theta_{18} &+ c_{11}~\theta_{-18}
  \end{array}
\end{equation}
The  $su(2)$ theta functions are related to the standard 
Jacobi theta function via
\begin{equation}  
\begin{array}{l}
  \theta_{m,24} ({\tau\over 2}, {z\over 6}) = \sum_{n\in\Z}
  q^{12(n+{m\over 48})^2} ~y^{4(n+{m\over 48})} = q^{({m\over 8})^2 /3}
  ~y^{m\over 12} ~\th_3(4z+{m\tau\over 2} |24\tau)
  \end{array}
\end{equation}
Under full spectral flow, these $su(2)$ theta functions are shifted
into one each other: 
\begin{equation}
  \begin{array}{l}
  \theta_{m,24}({\tau\over 2}, {z\over 6})
  \stackrel{z\to z+\tau}{\longrightarrow} q^{-1/3}
  y^{-2/3} ~\theta_{m+8,24}({\tau\over 2}, {z\over 6}),
  \end{array}
\end{equation}
so that the fifteen characters split into three 
groups which are cyclicly permuted: 
\begin{equation}
  \begin{split}
&A \stackrel{z\to z+\tau}{\longrightarrow} B \to C
  \to D \to E \to F \to A~,\\
&G \to H \to I \to G~,\\
&J \to K \to L \to M \to N \to O \to J,
  \end{split}
\end{equation}
where we have omitted the incrementing factors of $q^{-1/3}
y^{-2/3}$, etc.

To build the various orbits of the $4^3$ theory, we consider all
possible homogeneous polynomials of degree 3 in $A,B,...,O$, respecting
our usual rules.  Note the following powers for the $y$-expansions: 
  \begin{tabular}[t]{ll}
$A,D,H$ have powers of $y$ in $\Z$, & $J,M$ in $\Z+{1\over 6}$, \\
$B,E,I$ in $\Z-{1\over 3}$, & $K,N$ in $\Z-{1\over 6}$, \\
$C,F,G$ in $\Z+{1\over 3}$, & $L,O$ in $\Z+{1\over 2}$.
  \end{tabular}

The twenty-three possible orbits are:
\begin{equation*}
  \begin{array}{rl}
{\rm NS}_1 &= A^3 + B^3 +C^3 +D^3 + E^3 + F^3 \\
{\rm NS}_2 &= G^3 +H^3 +I^3 \\
{\rm NS}_3 &= (BC+EF)H+(CD+FA)I +(DE+AB)G \\
{\rm NS}_4 &= AO^2 +BJ^2 +CK^2 + DL^2 + EM^2 +FN^2 \\
{\rm NS}_5 &= (JL+MO)G + (KM+NJ)H +(LN+OK)I \\
{\rm NS}_6 &= AL^2+BM^2+CN^2+DO^2+EJ^2+FK^2 
  \end{array}
\end{equation*}
{\small
\begin{equation*}
  \begin{array}{ll}
{\rm NS}_7 = A^2D+B^2E+C^2F+D^2A+E^2B+F^2C & {\rm NS}_{16} = (L^2+O^2)H +(J^2+M^2)I +(K^2+N^2)G \\
{\rm NS}_8 = ABC+ BCD +CDE +DEF +EFA+FAB  & {\rm NS}_{17} = (A+D)LO + (B+E)JM+(C+F)KN\\
{\rm NS}_9 = AEC +BFD & {\rm NS}_{18} = HLO + IJM+ GKN  \\
{\rm NS}_{10} = GHI & {\rm NS}_{19} = AJK+BKL+CLM+DMN+ENO+FOJ \\
{\rm NS}_{11} = ADH+BEI+CFG & {\rm NS}_{20} = DJK+EKL+FLM+AMN+BNO+COJ \\
{\rm NS}_{12} = (A^2+D^2)H +(B^2+E^2)I +(C^2+F^2)G & {\rm NS}_{21} = (JK+MN)H+(KL+NO)I+(LM+OJ)G \\
{\rm NS}_{13} = (BF+EC)H+(AC+DF)I +(BD+AE)G & {\rm NS}_{22} = CJL+DKM+ELN+FMO+ANJ+BOK \\
{\rm NS}_{14} = (A+D)H^2 +(B+E)I^2+(C+F)G^2 & {\rm NS}_{23} = FJL+AKM+BLN+CMO+DNJ+EOK  \\
{\rm NS}_{15} = (C+F)HI+(A+D)IG+(B+E)GH & \\
  \end{array}
\end{equation*}
}
The coefficients $D_i$, see after (\ref{S_ij}), are thus
(1; 2,6,3,6,3; 3,6,18,36,12,3,6,3,6; 3,6,12,6,6,6,6,6).

\subsection{String Functions:}

Some explicit expressions for the string functions at level 2 are found in
\cite{KP-84}, pp. 219-220.  We use the notation \underline{$\eta_n$ for
$\eta(n\tau)$}:  
\begin{equation}
  \begin{split}
    c_{02} = {\eta_{12}^2 \over \eta^2 ~\eta_{6}}, \qquad \qquad
  c_{00} -c_{04} &= {1 \over \eta_{2}}, \qquad \qquad 
  c_{11} +c_{13} = {1 \over \eta_{1/2}}, \\
  c_{00} +c_{04} -2c_{02} + 2c_{20} -2c_{22} &= {\eta_{1/12}^2 \over
    \eta^2 ~\eta_{1/6}}
  \end{split}
\end{equation}
The behaviour under $S$ and $T$ transformation is also outlined by the
authors.   For example, $T$ transforms the third equation into
$$
c_{11} -c_{13} = {\eta_{1/2}~ \eta_{2} \over \eta^3},
$$
where we discarded $e^{i\pi/24}$ on both sides.  Similarly,
$T^6$ (ie $\tau \to \tau + 6$) transforms the last equation into
\begin{equation}
-\big(c_{00} +c_{04} +2c_{02} + 2c_{20} +2c_{22}\big) =
   -{\eta_{1/6}^5 \over \eta^2~ \eta_{1/12}^2 ~\eta_{1/3}^2}, \qquad
   {\rm since} \quad \eta_{1/2} \stackrel{T}{\longrightarrow}
{\eta^3 \over \eta_{1/2} ~\eta_{2} } ~e^{i\pi /24}. 
\end{equation}
Furthermore,
\begin{equation}
  \begin{array}{rl}
    c_{00} &\stackrel{S}{\longrightarrow} ~{1\over
    2\sqrt{6}}\sqrt{i\over\tau}\big( c_{00} +c_{04} +2 c_{02} +
    2c_{20} +2c_{22} + 2\sqrt{3}(c_{11} +c_{13}) \big) \\
    c_{04} &\stackrel{S}{\longrightarrow} ~{1\over
    2\sqrt{6}}\sqrt{i\over\tau}\big( c_{00} +c_{04} +2 c_{02} +
    2c_{20} +2c_{22} - 2\sqrt{3}(c_{11} +c_{13}) \big) \\
    c_{02} &\stackrel{S}{\longrightarrow} ~{1\over
    2\sqrt{6}}\sqrt{i\over\tau}\big( c_{00} +c_{04} -2 c_{02} +
    2c_{20} -2c_{22} \big) \\
    c_{20} &\stackrel{S}{\longrightarrow} ~{1\over
    2\sqrt{6}}\sqrt{i\over\tau}\big( 2c_{00} +2c_{04} +4c_{02} -
    2c_{20} -2c_{22} \big) \\
    c_{00} &\stackrel{S}{\longrightarrow} ~{1\over
    2\sqrt{6}}\sqrt{i\over\tau}\big( 2c_{00} +2c_{04} -4c_{02} -
    2c_{20} +2c_{22} \big),
  \end{array}
\end{equation}
and so 
\begin{equation}
  \begin{array}{rl}
    c_{00} +c_{04} \stackrel{S}{\longrightarrow} ~{1\over
    2\sqrt{6}}\sqrt{i\over\tau}\big( 2c_{00} +2c_{04} +4c_{02} +
    4c_{20} +4c_{22} \big) &= {1\over
    2\sqrt{6}}\sqrt{i\over\tau}\big( {\eta_{1/6}^5 \over
    \eta^2 ~\eta_{1/12}^2 ~\eta_{1/3}^2} \big)\\
    & \stackrel{S}{\longrightarrow} 
    {\eta_{6}^5 \over \eta^2 ~\eta_{12}^2 ~\eta_{3}^2 },
  \end{array}
\end{equation}
since $ \eta_{n} \stackrel{S}{\longrightarrow} \sqrt{-i\tau/n}
~\eta_{1/n}$ for rational $n$.  We thus find
\begin{equation*}
  \begin{split}
     c_{00} +c_{04} &= {\eta_{6}^5 \over \eta^2 ~\eta_{12}^2
     ~\eta_{3}^2 } = {\th_3(0|6\tau)\over \eta^2} \\
     c_{00} -c_{04} &= {1 \over \eta_{2}} =
     {\th_4(0|2\tau)\over \eta^2}\\
     c_{02} &=  {\eta_{12}^2 \over \eta^2 ~\eta_{6}} =
     {\th_2(0|6\tau)\over 2\eta^2}\\
  \end{split}
\end{equation*}
\begin{equation*}
  \begin{array}{rl}
     c_{20} +c_{22} &= {1\over 2\eta^2}\Big( {\eta_{1/6}^5 \over
     \eta_{1/12}^2 ~\eta_{1/3}^2} - \th_3(0|6\tau) 
     -\th_2(0|6\tau) \Big)
     = {1\over 2\eta^2}\Big(\th_3(0|\tau/6) - \th_3(0|3\tau/2) \Big) =
     {q^{1/12}\over \eta^2} \th_3({\tau\over 2}|{3\tau\over 2}) \\ 
     c_{20} -c_{22} &= {1\over 2\eta^2}\Big( {\eta_{1/12}^2 \over
     \eta_{1/6}} - \th_3(0|6\tau)  +\th_2(0|6\tau) \Big) 
     =  {1\over 2\eta^2}\Big(\th_4(0|\tau/6) - \th_4(0|3\tau/2) \Big) =
     -{q^{1/12}\over \eta^2} \th_4({\tau\over 2}|{3\tau\over 2}), 
  \end{array}
\end{equation*}
where we have used lemma \ref{easy1}, \ref{easy2}. Thus
\begin{equation}
  \begin{split}
c_{20} &= {q^{1/3}\over \eta^2} \th_3(2\tau|{6\tau}) 
       ={q^{1/12}\over \eta^2} \th_2(\tau|{6\tau})\\
c_{22} &= {q^{1/3}\over \eta^2} \th_2(2\tau|{6\tau}) 
       ={q^{1/12}\over \eta^2} \th_3(\tau|{6\tau}).
  \end{split}
\end{equation}

\subsection{Relation with $1^6$ theory} \label{sec:relation}

With the above values of the string functions, we shall show some
coincidences of the characters of $4^3$ theory with those of $1^6$
theory.  We first note:
\begin{equation}
  \begin{array}{rll}
\theta_m + \theta_{m+24} &= \sum q^{3(n+{m\over 24})^2} y^{2(n+{m\over 24})} 
   &= q^{({m\over 8})^2/3} y^{m\over 12} ~\th_3(2z+{m\tau\over 4}|6\tau) \\
\theta_m - \theta_{m+24} &= \sum (-1)^n q^{3(n+{m\over 24})^2} y^{2(n+{m\over 24})} 
   &= q^{({m\over 8})^2/3} y^{m\over 12} ~\th_4(2z+{m\tau\over 4}|6\tau) \\
\theta_m + \theta_{m+12} + \theta_{m+24} + \theta_{m+36} 
&=  \sum q^{{3\over 4}(n+{m\over 12})^2} y^{n+{m\over 12}} 
&= q^{({m\over 8})^2/3} y^{m\over 12} ~\th_3(z+{m\tau\over
  8})|{3\tau\over 2}) \\ 
\theta_m - \theta_{m+12} + \theta_{m+24} - \theta_{m+36} 
&=  \sum (-1)^n q^{{3\over 4}(n+{m\over 12})^2} y^{n+{m\over 12}} 
&= q^{({m\over 8})^2/3} y^{m\over 12} ~\th_4(z+{m\tau\over
  8})|{3\tau\over 2})\\
  \end{array}
\end{equation}
\begin{equation*}
  \begin{array}{rl}
\theta_m + \theta_{m+12} - \theta_{m+24} - \theta_{m+36} 
&=  \sum \delta_n ~q^{{3\over 4}(n+{m\over 12})^2} y^{n+{m\over 12}},
\qquad \delta_n :=+,+,-,- \textrm{ for } n\equiv 0,1,2,3
\textrm{ mod } 4\\
\textrm{ for }m\equiv 2 \textrm{ mod } 8: &= \sum (q ~e^{2\pi
   i})^{{3\over 4}(n+{m\over 12})^2} y^{n+{m\over 12}} \qquad\\
&= q^{({m\over 8})^2/3 + 3/16} y^{m\over 12}~{\eta_3\over\eta_6~\eta_{3/2}}
~\th_3(z +({3\over 4}+{m\over 8})\tau|3\tau)
~\th_3(z +({3\over 4}-{m\over 8})\tau|3\tau) \\
\textrm{ and for} ~m\equiv -2 \textrm{ mod } 8: &=\sum (-1)^n (q ~e^{2\pi
   i})^{{3\over 4}(n+{m\over 12})^2} y^{n+{m\over 12}} \qquad \\
&= q^{({m\over 8})^2/3 + 3/16} y^{m\over 12}~{\eta_3\over\eta_6~\eta_{3/2}}
~\th_4(z +({3\over 4}+{m\over 8})\tau|3\tau) 
~\th_4(z +({3\over 4}-{m\over 8})\tau|3\tau) 
  \end{array}
\end{equation*}
The trick for the last formula is the same as in the special case
leading to (\ref{++--}).   

Thus, using lemma \ref{easy4}:
\begin{equation} \label{1^6-4^3}
  \begin{array}{rll}
A+D &= 2 c_{02} ~\th_2(2z|6\tau) +(c_{00}+c_{04}) ~\th_3(2z|6\tau)
= {1\over \eta^2} \th_3(z|3\tau)^2 &= A^2{ }_{[1^6 ~{\rm theory}]} \\
B+E &= {q^{1/3} y^{2/3}\over \eta^2} \th_3 (z+\tau|3\tau)^2 &= B^2{ }_{[1^6 ~{\rm theory}]} \\
C+F &= {q^{4/3} y^{4/3}\over \eta^2} \th_3 (z+2\tau|3\tau)^2 &= C^2{ }_{[1^6 ~{\rm theory}]} \\
G &= c_{22} ~q^{1\over 12} y^{1\over 3} ~\th_3(2z+\tau|6\tau) + 
 c_{20} ~q^{1\over 12} y^{1\over 3} ~\th_2(2z+\tau|6\tau) & \\
&= {q^{1/6} y^{1/3} \over \eta^2} ~\th_3(z|3\tau)~\th_3(z+\tau|3\tau) &= AB_{[1^6 ~{\rm theory}]} \\
H &= {q^{5/6} y \over \eta^2} ~\th_3(z+\tau|3\tau)~\th_3(z+2\tau|3\tau) &= BC_{[1^6 ~{\rm theory}]} \\
I &= {q^{2/3} y^{2/3} \over \eta^2} ~\th_3(z|3\tau)~\th_3(z+2\tau|3\tau) &= AC_{[1^6 ~{\rm theory}]},
  \end{array}
\end{equation}
where the rhs's are taken from (\ref{1^6-char}).  In particular, this
implies $(A+D)H=GI$, etc, and for the orbits:
$$
\begin{array}{r|l}
4^3 ~{\rm theory}  & 1^6 ~{\rm theory} \\ \hline
{\rm NS}_1 + 3~{\rm NS}_7 &= {\rm NS}_1\\
{\rm NS}_2 &= {\rm NS}_2\\
{\rm NS}_3 + {\rm NS}_{13} & = {\rm NS}_2 \\
{1\over 3}~{\rm NS}_{14} ={\rm NS}_{10} ={\rm NS}_8+ {\rm NS}_9 &= {\rm NS}_3\\
{\rm NS}_{15} = {\rm NS}_{12}+2~{\rm NS}_{11} &= {\rm NS}_4.  
\end{array}
$$
These relations will be useful for the study of
Gepner models with mixed levels, in particular the $1^4 4$ theory.

\subsection{Characters at $z={1+\tau\over 2}$:}

We now study the orbits at the special value of
\underline{$z={1+\tau\over 2}$}.  Note first that at this value,
$\theta_m =\theta_{m,24}({\tau\over 2}, {1+\tau\over 12}) = e^{2\pi i
  {m\over 24}} q^{({m^2\over 8} +m)/24}~ \th_3(({m\over 2}+2)\tau
|24\tau)$, as well as:
\begin{equation}
  \begin{array}{rl}
\theta_8=\theta_{-16}, \qquad \qquad \theta_{-22} &= -\theta_{14},
\qquad\qquad \theta_{-10}=-\theta_{2} \\
\theta_m + \theta_{m+12} + \theta_{m+24} + \theta_{m+36} 
&= e^{2\pi i {m\over 24}} q^{({m^2\over 8} +m)/24}~ \th_4(({m\over
  8}+\half)\tau|{3\tau\over 2}) \\ 
\theta_m + \theta_{m+12} -\theta_{m+24} - \theta_{m+36} &=  e^{2\pi i
  {m\over 24}} \sum \delta_n ~q^{{3\over 4}(n+{m\over 12})^2 +{1\over
  2}(n+{m\over 12})},
\end{array}
\end{equation}
where $\delta_n = +,+,-,-$ for $n\equiv 0,1,2,3$ mod 4.  Without
$\delta_n$, we would recover the sum of the four theta functions with
only $+$ signs.  Note that $\delta_n$ can be removed if we replace
$q^{1/2}$ with $-q^{1/2}$ in the sum and additionally multiply the
sum by some root of unity, since $(-1)^{{3\over
    2}(n+{m\over 12})^2 + n+{m\over 12} } =i^{{m^2\over 48}
  +{m\over 6}} ~i^{3n^2 + (2+{m\over 2}) n}$.  For $m=6$ or $-2$,
we recover $e^{-2\pi i/16} \delta_n$. Thus we obtain the last line
from the sum with only $+$ signs by inserting $-q^{1/2}$ in the
latter's result.   For $m=6$ and $-2$:
\begin{equation}
  \begin{array}{rl}
    \theta_6 + \theta_{18} + \theta_{-18} + \theta_{-6} &= -i
    q^{-{1\over 16}}~ \th_4({-{\tau\over 4}}|{3\tau\over 2})
    = -i q^{-{1\over 12}}~ \eta({\tau\over 2}) \\
\theta_{-2} + \theta_{10} + \theta_{22} + \theta_{-14} &= 
    q^{-{1\over 12}}~ \eta({\tau\over 2}) 
  \end{array}
\end{equation}
On the rhs, replacing $q^{1/2}$ with $-q^{1/2}$ in $q^{-{1\over 12}}~
\eta(\tau/2)$ yields:
\begin{equation}
  q^{-{1\over 16}} \prod(1-q^n)(1-q^{n-1/2}) \longrightarrow e^{-2\pi
  i/16}~ q^{-{1\over 16}} \prod(1-q^n)(1+q^{n-1/2}) = e^{-2\pi
  i/16}~ q^{-{1\over 12}} {\eta^3\over \eta(\tau/2) \eta(2\tau)}
\end{equation}
Thus:
\begin{equation} \label{++--}
  \begin{array}{rl}
    \theta_6 + \theta_{18} - \theta_{-18} - \theta_{-6} &= -i
    ~q^{-{1\over 12}} {\eta^3\over \eta(\tau/2) \eta(2\tau)}\\
\theta_{-2} + \theta_{10} - \theta_{22} - \theta_{-14} &= 
    q^{-{1\over 12}} {\eta^3\over \eta(\tau/2) \eta(2\tau)}.
  \end{array}
\end{equation}
This will be useful for the characters $K,L,N,O$.

With lemmas \ref{eta-square-1},\ref{eta-square-2} and
\ref{eta-square-3} our characters at $z={1+\tau\over 2}$ reduce to
\begin{equation*}
  \begin{array}{rl}
A &= {1\over 2\eta^2}\Big[ \th_2(0|6\tau) (\theta_{-12} +
 \theta_{12}) + \th_3(0|6\tau) (\theta_{0} + \theta_{24}) + 
  \th_4(0|2\tau) (\theta_{0} -\theta_{24})\Big]\\ 
  &= {1\over 2\eta^2}\Big[ \th_2(0|6\tau)
 (-q^{1/4})\th_3(2\tau|6\tau)
 +\th_3(0|6\tau)~\th_3(\tau|6\tau)
 +\underbrace{\th_4(0|2\tau)~\th_4(\tau|6\tau)}_{q^{-1/12}~\eta^2} \Big] = q^{-1/12} \\
B &= C= D = E = 0\\
F &= e^{-2\pi i/3} ~q^{-1/12}\\
G &= {q^{1/12}\over \eta^2} \th_3(\tau|6\tau) ~(\theta_{-20} +
 \theta_{4}) + {q^{1/3}\over \eta^2} \th_3(2\tau|6\tau) ~(\theta_{-8} +
 \theta_{16}) = 0 \\
H &= 0 \\
I &=  e^{-2\pi i/6} ~q^{-1/12}\\
  \end{array}
\end{equation*}
\begin{equation*}
  \begin{array}{rl}
J &= c_{13} (\theta_{-22} +\theta_{14}) +c_{11} (\theta_{-10}
 +\theta_{2}) =0 \\
K &= \half (c_{11}+c_{13})(\theta_{-14} +\theta_{-2}
 +\theta_{10}+ \theta_{22}) + \half
 (c_{11}-c_{13})(-\theta_{-14} +\theta_{-2} +\theta_{10}
 -\theta_{22}) = e^{-2\pi i/12} ~q^{-1/12} \\
L &= M = N = 0 \\
O &= -i ~q^{-1/12}.
  \end{array}
\end{equation*}

Plugging these values into the orbits NS$_i$ yields NS$_1 = 2
q^{-1/4}$, NS$_i = -q^{-1/4}$ ($i=2,\dots,6$), while the remaining
NS$_j$ vanish ($j=7,\dots,23$). We thus recognize from
(\ref{NS-orbits}) the graviton, massless and massive orbits
respectively.  Also, the value of the elliptic genus at $z=0$ is $
\Phi (0) =\sum_{i=1}^d D_i ~|{\rm R}'_i(0)|^2 = \sum_{i=1}^d D_i
~|-q^{1/4} {\rm NS}_i({1+\tau\over 2})|^2 = \sum_{i=1}^d D_i I_i^2 =
4D_1 + D_2 +\dots +D_6 = 24$, which is the correct coefficient for a
$K3$ model (\ref{ell-genus}).

\subsection{Characters at $z=0$:}

In order to compute the functions $F_i$ or the Dirac genus, we set
$z=0$, in which case 
\begin{equation*}
  \begin{array}{rl}
\theta_m &= \theta_{m,24}({\tau\over 2},0) = \sum
    q^{12(n+{m\over 48})^2} =q^{({m\over
        8})^2/3}~\th_3({m\tau\over 2} |24 \tau) =\theta_{-m}\\ 
\theta_m +\theta_{m+12} -\theta_{m+24} -\theta_{m+36}
    &= \sum \delta_n q^{{3\over 4}(n+{m\over 12})^2} = \sum
    \delta_n q^{{3\over 4}(n-1-{m\over 12})^2} \\
  ({\rm for }~ m=-10,-2:)  &= \sum \delta_n q^{{3\over 4}(n-{1\over 6})^2}
    = e^{-2\pi i({m\over 8})^2/3} \sum (-1)^n (-q^{1/2})^{{3\over
    2}(n-{1\over 6})^2} = {\eta^3 \over \eta_{1/2}~\eta_2},
  \end{array} 
\end{equation*} 
where $\delta_n := +,+,-,- $ and the last line is obtained by
replacing $q^{1/2}$ by $q^{-1/2}$ in $\eta(\tau/2)$ (same trick as
earlier).

With lemma \ref{easy4}, the characters at $z=0$ take the following
values:
\begin{equation}
  \begin{array}{rl}
A &= {1\over 2\eta^2}\th_3(0|3\tau)^2 +
     {1\over 2\eta^2} \th_4(0|2\tau)~\th_4(0|6\tau) =: A_1 +A_2 \\
B = F &= {q^{1/3}\over 2\eta^2}~\th_3(\tau|3\tau)^2  +
     {q^{1/3}\over 2\eta^2}~\th_4(0|2\tau)~\th_4(2\tau|6\tau) =: B_1 +B_2  \\
C = E &= B_1 -B_2\\
D &= A_1 - A_2 \\
  \end{array}
\end{equation}
\begin{equation}
  \begin{array}{rl}
G =I &= {q^{1/6}\over \eta^2}~\th_3(0|3\tau)~\th_3(\tau|3\tau)\\
H &= {q^{1/3}\over \eta^2}~\th_3(\tau|3\tau)^2  =2 B_1\\
J=K &= {q^{1/48} \over 2\eta_{1/2}}~\th_3({\tau\over 4}|{3\tau\over 2}) + \half =: J_1 + \half  \\
L =O &= {1\over 2\eta_{1/2}}\th_2(0|{3\tau\over 2})  \\
M =N &= J_1-\half.
  \end{array}
\end{equation}
We did not succeed in factorizing them, as we did for the $1^6$ and
$2^4$ theories.  The corresponding values for the orbits are not
particularly enlightening.  We only note the following coincidence:
NS$_{22}$ = NS$_{23} = 2A_1(J_1^2-{1\over 4}) +4 L J_1 B_1$.  Due to
\ref{NS-orbits}, this equality holds in general (not only at $z=0$)
since $F_{22} =F_{23}$.

Thus we shall also refrain from giving here horrendous expressions
(unfactorized) for the functions $F_i$ and the Dirac index.  But they
can be easily written down on the basis of the above information.
For instance, proving the value of the Dirac index (\ref{dirac-index})
boils down to verifying its coefficient at $z=0$: 
$$
\begin{array}{rl}
\sum_{i=1}^d & D_i ~{\rm NS}_i(0) ~I_i = -2 ~{\rm NS}_1 +2 ~{\rm
  NS}_2 + 6~{\rm NS}_3 +3~ {\rm NS}_4 +6~ {\rm NS}_5 +3~ {\rm NS}_6 \\
&=32 B_1^3 -4A_1^3 -12 A_1 A_2^2 -48 B_1 B_2^2 +4G^3 +24 G (A_1B_1 +
A_2B_2 +J_1 L) +24 B_1 J_1^2 +12 A_1 L^2 \\
&\stackrel{!}{=}  2 {\th_2^4-\th_4^4 \over\eta^6}~\th_3^2,
\end{array}
$$
which is an arduous manipulation with theta functions identities (left
to the reader).

\subsection{Lemmas and arithmetic results:}

\begin{lem}\label{easy1}
  \begin{equation}
    \begin{split}
      \th_3(z|\tau) &= \th_3(2z|4\tau) + \th_2(2z|4\tau) \\ 
      \th_4(z|\tau) &= \th_3(2z|4\tau) - \th_2(2z|4\tau)
    \end{split}
  \end{equation}
\end{lem}
\begin{proof}
  Directly from Fourier expansion.
\end{proof}

\begin{lem}\label{easy2}
  \begin{equation}
    \begin{split}
      \th_2(0|\tau) &= \th_2(0|9\tau) + 2q^{1/2}~\th_2(3\tau|9\tau) \\
      \th_3(0|\tau) &= \th_3(0|9\tau) + 2q^{1/2}~\th_3(3\tau|9\tau) \\
      \th_4(0|\tau) &= \th_4(0|9\tau) - 2q^{1/2}~\th_4(3\tau|9\tau) \\
    \end{split}
  \end{equation}
\end{lem}
\begin{proof}
Idem. For instance, the middle line goes like 
\begin{equation*}
  {\rm lhs} = \sum q^{(3n)^2/2} +\sum q^{(3n+1)^2/2} +\sum q^{(3n+2)^2/2} 
=  \sum q^{(3n)^2/2} + 2 \sum q^{(3n+1)^2/2} = {\rm rhs}.
\end{equation*}  
\end{proof}

\begin{lem}\label{easy3}
  \begin{equation}
    \begin{array}{lll}
      \th_3({1\over 3}|{2\tau\over 3}) &= \th_3(0|6\tau)
    -q^{1/3}~\th_3(2\tau|6\tau) &= \th_3(0|6\tau) -q^{1/12}~\th_3(\tau|6\tau)\\
      \th_2({1\over 3}|{2\tau\over 3}) &= -\th_2(0|6\tau)
   +q^{1/3}~\th_2(2\tau|6\tau) &= -\th_2(0|6\tau) +q^{1/12}~\th_3(\tau|6\tau)\\
    \end{array}
  \end{equation}
\end{lem}
\begin{proof}
Idem.  For instance, the first line:
\begin{equation*}
  {\rm lhs} = \sum q^{n^2 /3} e^{2\pi in/3} 
= \sum q^{3n^2 /3} + (e^{2\pi i/3} + e^{-2\pi i/3})
  \sum q^{3(n+{1\over 3})^2 /3} =  {\rm rhs}.
\end{equation*}
\end{proof}

\begin{lem} \label{easy4}
  \begin{equation}
    \begin{array}{ll}
       \th_3(z|\tau)~\th_3(z'|\tau) + \th_2(z|\tau)~\th_2(z'|\tau)&=
       \th_3({z+z'\over 2}|{\tau\over 2})~\th_3({z-z'\over 2}|{\tau\over 2}) \\
       \th_3(z|\tau)~\th_3(z'|\tau) - \th_2(z|\tau)~\th_2(z'|\tau)&=
       \th_4({z+z'\over 2}|{\tau\over 2})~\th_4({z-z'\over 2}|{\tau\over 2})
    \end{array}
  \end{equation}
\end{lem}
\begin{proof}
Idem.  For instance, the first line:
\begin{equation*}
  {\rm lhs} = \sum_{\Z^2 \cup (\Z+\half)^2} q^{(m^2 +n^2)/2} y^m {y'}^n = \sum_{\Z^2} q^{(k^2 +l^2)/4} y^{(k+l)/2} {y'}^{(k+l)/2} =  {\rm rhs},
\end{equation*}
where we made the substitution $k=m+n, ~~l=m-n$.  

For the second line, one just needs to introduce $(-1)^{2n}$ and
$(-1)^{k-l}$ in the two sums respectively.
\end{proof}

\begin{lem} \label{eta-square-1}
  \begin{equation}
    \begin{array}{lll}
       \th_3(0|6\tau)~\th_3(\tau|6\tau) &-\th_2(0|6\tau)~\th_2(\tau|6\tau)
         &= q^{-1/12}~\eta^2 \\
  \th_3(0|{\tau\over 6})~\th_3({1\over 6}|{\tau\over 6})
&- \th_4(0|{\tau\over 6})~\th_4({1\over 6}|{\tau\over 6}) &= 6\eta^2 \\ 
  \th_3(0|{2\tau\over 3})~\th_2({1\over 3}|{2\tau\over 3})
&+\th_2(0|{2\tau\over 3})~\th_3({1\over 3}|{2\tau\over 3})&= 3\eta^2 \\ 
  \th_3(0|{3\tau\over 2})~\th_4(\tau|{3\tau\over 2})
& -\th_4(0|{3\tau\over 2})~\th_3(\tau|{3\tau\over 2}) &= -2q^{-1/3}~\eta^2 \\ 
    \end{array}
  \end{equation}
\end{lem}
\begin{proof}
  The first line is just lemma \ref{easy4} with $z=0$ and $(z'|\tau)$
  replaced by $(1+\tau|6\tau)$.  The second line is obtained from the
  first by $S$ transformation, ie $\tau \to -1/\tau$.  The third line
  is obtained by rewriting the second line as $ab+cd= (a+c)(b+d)/2
  +(a-c)(b-d)/2$ and using lemma \ref{easy1}.  The fourth line is
  again an $S$ transformation of the previous line. Applying the
  rewriting trick on this last line, we would of course fall back on
  the first line.
\end{proof}

We note that the first line in Fourier series gives us an interesting
formula: $ \eta^2 = \sum (-1)^n q^{3(n^2+m^2)}$ where $(m,n)\in
(\half,{1\over 3}) \cup (0,{1\over 6}) + \Z^2$. This is to be
compared with the previous formula \ref{eta-square-0}.  The crucial
difference is that we presently have a positive definite quadratic
form in the exponent of $q$, whereas previously the form was
indefinite (this accounts for the extra constraint on $x,y$ there). 

In the first line, use lemma \ref{easy2} to replace $2
q^{1/12}~\th_3(\tau|6\tau)$ by $ \th_2(0|{2\tau\over 3})
-\th_2(0|6\tau)$ and similarly for $2q^{1/12}~\th_2(\tau|6\tau)$, and
obtain:
\begin{lem} \label{eta-square-2}
  \begin{equation}
    \begin{array}{ll}
  \th_2(0|{2\tau\over 3})~\th_3(0|6\tau) -\th_3(0|{2\tau\over
  3})~\th_2(0|6\tau) &= 2 \eta^2 \\  
  \th_4(0|{3\tau\over 2})~\th_3(0|{\tau\over 6}) -\th_3(0|{3\tau\over
  2})~\th_4(0|{\tau\over 6}) &= 4 \eta^2,
    \end{array}
  \end{equation}
\end{lem}
where an $S$ transformation connects the two lines.  Unlike in the
previous lemma, performing the rewriting trick will not give two more
variants; here this trick is just equivalent to the $S$ transformation
itself.

Now, in the third line of lemma \ref{eta-square-1}, use lemma
\ref{easy3} to replace
$\th_2({1\over 3}|{2\tau\over 3})$ by $-\th_2(0|6\tau) +
q^{1/3}~\th_2(2\tau|6\tau)$ and similarly for $\th_3({1\over
  3}|{2\tau\over 3})$, and obtain the first line of

\begin{lem} \label{eta-square-3}
  \begin{equation}
    \begin{array}{lll}
\th_3(0|{2\tau\over 3})~\th_3(\tau|6\tau) &-\th_2(0|{2\tau\over
  3})~\th_2(\tau|6\tau) &= q^{-1/12}~\eta^2 \\
\th_3(0|{3\tau\over 2})~\th_3({1\over 3}|{\tau\over 6})
 & -\th_4(0|{3\tau\over 2})~\th_4({1\over 3}|{\tau\over 6}) &= 2\eta^2 \\  
\th_3(0|6\tau)~\th_2({2\over 3}|{2\tau\over 3})
 &+ \th_3(0|6\tau)~\th_3({2\over 3}|{2\tau\over 3})&= \eta^2 \\ 
\th_4(0|{\tau\over 6})~\th_3(\tau|{3\tau\over 2})
&-\th_3(0|{\tau\over 6})~\th_4(\tau|{3\tau\over 2}) &=q^{-1/3}~\eta^2. 
    \end{array}
  \end{equation}
\end{lem}
\begin{proof}
  Again, the successive lines are obtained by $S$ transformation,
the rewriting trick, and $S$ transformation.
\end{proof}

\section{Computations in Mixed Theories}

We have met with success the construction of N=4 characters in {\em
  pure} Gepner models like $1^6, 2^4, 4^3$ theories.  Other Gepner
models for the N=4 SCFT on $K3$ are {\em mixed} tensor products of N=2
theories, like $1^3 2^2$, $1^4 4$, $1^2 4^2$, $2 ~6^2$, $1 ~2^2 4$,
$\dots$.  All these products $k_1^{n_1}\dots k_j^{n_j}$ are formed
with the requirement that the central charge equal six: $c=\sum n_i
{3k_i\over k_i + 2} = 6$.  We shall investigate the first three cases
of such theories with mixed levels $k_i$ and see that they do not
necessarily share the previous structure characteristic of {\em pure}
theories.  Specifically, the notion of gravitational, massless and
massive orbits, with values at $z={1+\tau \over 2}$ equal to
$2q^{-1/4}, -q^{-1/4}$ and 0 respectively -- expected from
(\ref{NS-orbits}), only applies to $K3$ models like the $1^4 4$ or
$1^2 4^2$ theories below.  The other CY twofold, the complex torus,
gives a model whose orbits all vanish at $z={1+\tau \over 2}$,
yielding a zero Euler characteristic as in the $1^3 2^2$ theory below. 

\subsection{Computations in $1^3 2^2$ Theory}

The NS$_i$ orbits are tensor products of three orbits from the $k=1$
theory (section \ref{sec:1^6-theory}) and two orbits from the $k=2$
theory (section \ref{sec:2^4-theory}).  The former theory has
characters $A,B,C$ (with powers of $y$ in $\Z+0,{1\over 3},{2\over 3}$
resp.), while the latter's characters we denote by $\bar{A}, \bar{B},
\bar{C}, \bar{D}, \bar{E}, \bar{F}$ (with powers of $y$ in $\Z+0,
\half, 0, \half, -{1\over 4}, {1\over 4}$ resp.).  In order to have
only integer powers of $y$ and cyclic permutation in the orbits, the
only possible combinations are products of $ABC$, $A^3+B^3+C^3$ with
$\bar{A} \bar{C} +\bar{B} \bar{D}$, $\bar{E} \bar{F}$, $\bar{A}^2
+\bar{C}^2 +\bar{B}^2 +\bar{D}^2 $.  Each of these products vanishes
at $ z={1+\tau \over 2}$, so all orbits are massive and the Dirac
index vanishes.  This is to be expected as both the $1^3$ and the $2^2$
theories are toroidal models (with complex and K\"ahler moduli
$\tau=\rho=e^{2\pi i/3}$ and $\tau=\rho=i$ resp.), hence so is their
tensor product.  That is, the target space of the sigma model is not
$K3$ but a complex two-torus.

\subsection{Computations in $1^4 4$ Theory}

We denote the characters of the $k=4$ theory by a bar over the
letters.  Our usual two rules (integer powers of $y$ and
invariance under cyclic permutation) restrict the orbits to be of the
following form:
\begin{equation*}
  \begin{array}{lll}
{\rm NS}_1 =  A^4 (\bar{A}+\bar{D}) + B^4 (\bar{B}+\bar{E}) +C^4
    (\bar{C}+\bar{F}) & \qquad & 
{\rm NS}_5 = ABC (A\bar{H}+B\bar{I}+C\bar{G}) \\ 
{\rm NS}_2 =  A^3 B (\bar{B} +\bar{E}) +B^3 C (\bar{C} +\bar{F}) +C^3
    A(\bar{A}+\bar{D}) & \qquad &  
{\rm NS}_6 = A^4 \bar{H} + B^4 \bar{I} +C^4 \bar{G} \\
{\rm NS}_3 = A^3 C (\bar{C} +\bar{F}) +B^3 A (\bar{D} +\bar{A}) +C^3
    B(\bar{E}+\bar{B}) & \qquad &  
{\rm NS}_7 = A^3 B \bar{I} +B^3 C \bar{G} +C^3 A \bar{H} \\
{\rm NS}_4 = A^2 B^2 \bar{G} + B^2 C^2 \bar{H} + C^2 A^2 \bar{I} & \qquad & 
{\rm NS}_8 = A^3 C \bar{G} +B^3 A \bar{H} +C^3 B \bar{I}
  \end{array}
\end{equation*}
Due to the relations between the $4^3$ characters and the $1^6$
characters established in (\ref{1^6-4^3}), the orbits 2,3,4 are equal,
and so are the orbits 6,7,8. Thus we obtain consecutively the orbits 
NS$_1$,  NS$_2$,  3 NS$_3$  and NS$_4$ of $1^6$ theory
(\ref{1^6-orbits}), which proves the equivalence of both models!

The coefficients $D_i=S_{1,i}/S_{i,1}$ defined after (\ref{S_ij}) are
(1; 4,4,12; 24; 2,8,8), where $S_{i,1}=(6; 6,6,3; 3; 3,3,3)$ are the numbers
of terms in each orbit and $S_{1,i}=(1; 4,4,6; 12; 1,4,4)$ is $S_{1,1}$
times the number of permutations of the factors in any term of orbit
NS$_i$ (look only at the $1^4$ factors). Thus for the Dirac index, we
have correctly $-2$ NS$_1$ + (4+4+12) NS$_2$, as in
(\ref{1^6-dirac-index}).

\subsection{Computations in $1^2 4^2$ Theory}

This time, we are even allowed to include the characters $J,K,..,O$
from the $k=4$ theory.  The orbits take the form
{\small
\begin{equation*}
  \begin{array}{ll}
{\rm NS}_1 = A^2 (\bar{A}^2 +\bar{D}^2) +B^2(\bar{B}^2 +\bar{E}^2)
+C^2(\bar{C}^2 +\bar{F}^2) \\
{\rm NS}_2 = A^2 (\bar{L}^2 +\bar{O}^2) +B^2(\bar{J}^2 +\bar{M}^2)
+C^2(\bar{K}^2 +\bar{N}^2) \\
{\rm NS}_3 = A^2 (\bar{B} +\bar{E})\bar{G} +B^2
(\bar{C} +\bar{F})\bar{H} +C^2 (\bar{D} +\bar{A})\bar{I} \\
{\rm NS}_4 =  A^2 (\bar{C} +\bar{F})\bar{I} +B^2 (\bar{D} +\bar{A})\bar{G} +C^2
(\bar{E}+\bar{B})\bar{H} \\
{\rm NS}_5 = AB \bar{G}^2 +BC \bar{H}^2 +CA \bar{I}^2 \\
{\rm NS}_6 = AB (\bar{A}\bar{B}+\bar{D}\bar{E})
+BC(\bar{B}\bar{C}+\bar{E}\bar{F}) +CA(\bar{C}\bar{D}+\bar{F}\bar{A}) \\
{\rm NS}_7 = AB (\bar{L}\bar{J}+\bar{O}\bar{M})
+BC(\bar{M}\bar{K}+\bar{J}\bar{N}) +CA(\bar{N}\bar{L}+\bar{K}\bar{O}) \\
  \end{array}
\end{equation*}
\begin{equation*}
  \begin{array}{l|l}
{\rm NS}_8 = A^2 \bar{H}^2 +B^2 \bar{I}^2 +C^2 \bar{G}^2 
& {\rm NS}_{17} = A^2 \bar{A}\bar{D} +B^2 \bar{B}\bar{E} +C^2 \bar{C}\bar{F}\\
{\rm NS}_9 = AB (\bar{C}+\bar{F})\bar{G} +BC (\bar{A}+\bar{D})\bar{H} + CA
 (\bar{B}+\bar{E})\bar{I} 
& {\rm NS}_{18} = A^2 \bar{L}\bar{O} +B^2 \bar{J}\bar{M} +C^2 \bar{K}\bar{N}\\ 
{\rm NS}_{10} = AB \bar{H}\bar{I} +BC \bar{I}\bar{G} +CA \bar{G}\bar{H}
& {\rm NS}_{19} =  AB \bar{C}\bar{F} +BC \bar{D}\bar{A} +CA \bar{E}\bar{B} \\
{\rm NS}_{11} = A^2 (\bar{B}\bar{C} +\bar{E}\bar{F}) +B^2(\bar{C}\bar{D}
 +\bar{F}\bar{A}) +C^2(\bar{D}\bar{E} +\bar{A}\bar{B})
& {\rm NS}_{20} = AB (\bar{A}\bar{E}+\bar{D}\bar{B}) +BC(\bar{B}\bar{F}+\bar{E}\bar{C}) +CA(\bar{C}\bar{A}+\bar{F}\bar{D}) \\
{\rm NS}_{12} =  A^2 (\bar{B}\bar{F} +\bar{E}\bar{C}) +B^2(\bar{C}\bar{A}
 +\bar{F}\bar{D}) +C^2(\bar{D}\bar{B} +\bar{A}\bar{E})
& {\rm NS}_{21} = AB (\bar{L}\bar{M}+\bar{O}\bar{J})
+BC(\bar{M}\bar{N}+\bar{J}\bar{K}) +CA(\bar{N}\bar{O}+\bar{K}\bar{L}) \\
{\rm NS}_{13} = A^2 \bar{I}\bar{G} +B^2 \bar{G}\bar{H} +C^2 \bar{H}\bar{I}
& {\rm NS}_{22} =  A^2 (\bar{J}\bar{N} +\bar{M}\bar{K}) +B^2(\bar{K}\bar{O} +\bar{N}\bar{L}) +C^2(\bar{L}\bar{J} +\bar{O}\bar{M})  \\
{\rm NS}_{14} = A^2 (\bar{A} +\bar{D})\bar{H} +B^2 (\bar{B} +\bar{E})\bar{I} +C^2 (\bar{C} +\bar{F})\bar{G}
& {\rm NS}_{23} = A^2 (\bar{J}\bar{K} +\bar{M}\bar{N}) +B^2(\bar{K}\bar{L} +\bar{N}\bar{O}) +C^2(\bar{L}\bar{M} +\bar{O}\bar{J}) \\
{\rm NS}_{15} = AB (\bar{B}+\bar{E})\bar{H} +BC
(\bar{C}+\bar{F})\bar{I} +CA (\bar{A}+\bar{D})\bar{G}
& {\rm NS}_{24} = AB \bar{K}\bar{N} +BC \bar{L}\bar{O} +CA \bar{J}\bar{M} \\
{\rm NS}_{16} = AB (\bar{A}+\bar{D})\bar{I} +BC (\bar{B}+\bar{E})\bar{G} +CA (\bar{C}+\bar{F})\bar{H}
& {\rm NS}_{25} = AB (\bar{K}^2+\bar{N}^2) +BC (\bar{L}^2+\bar{O}^2) + CA
 (\bar{J}^2+\bar{N}^2) \\
& {\rm NS}_{26} = AB (\bar{C}^2+\bar{F}^2) +BC (\bar{D}^2+\bar{A}^2) + CA
 (\bar{E}^2+\bar{B}^2)
  \end{array}
\end{equation*}
}

At $ z={1+\tau \over 2}$, the first two orbits give $2q^{-1/4}$ and
$-2q^{-1/4}$ resp., while the next five orbits give $q^{-1/4}$; the
other orbits all give 0.  This embarrassing second orbit prevents us
to classify it as either a graviton, massless or massive orbit.  The
coefficients $D_i$ are (1,1,2,2,4,4,4; 2,4,8,2,2,4,2,4,4;
4,4,4,4,4,2,2,8,2,2).  So the value of the elliptic genus at $z=0$ is
$ \Phi (0) =\sum_{i=1}^d D_i ~|{\rm R}'_i(0)|^2 = \sum_{i=1}^d D_i
I_i^2 = 4D_1 + 4 D_2 +D_3+\dots +D_7 = 24$, so this is a $K3$ model
with the appropriate Euler character.

Due to the relations between the $4^3$ characters and the $1^6$
characters established in (\ref{1^6-4^3}), we find the following
relations between the orbits:
$$
\begin{array}{r|l || r|l || r|l}
1^2 4^2 ~{\rm theory} & 4^3 ~{\rm theory} & 1^2 4^2 ~{\rm theory} & 4^3~{\rm theory} & 1^2 4^2 ~{\rm th.} & 4^3~{\rm theory} \\
\hline
{\rm NS}_1 & {\rm NS}_1 + {\rm NS}_7 &   {\rm NS}_8= {\rm NS}_9= {\rm NS}_{10} & {\rm NS}_{14} =3~{\rm NS}_{10} & {\rm NS}_{18} & {\rm NS}_{17} \\

{\rm NS}_2 & {\rm NS}_4 + {\rm NS}_6 & {\rm NS}_{11} & 2~{\rm NS}_8   & {\rm NS}_{19} & {\rm NS}_{11} \\

{\rm NS}_3= {\rm NS}_4= {\rm NS}_5 & {\rm NS}_3 + {\rm NS}_{13} & {\rm NS}_{12} &
{\rm NS}_8 + 3~{\rm NS}_9 & {\rm NS}_{20} & {\rm NS}_{13} \\

{\rm NS}_6 & {\rm NS}_3 & {\rm NS}_{13}= {\rm NS}_{14}= {\rm NS}_{15}
& {\rm NS}_{15} = {\rm NS}_{12} + 2~{\rm NS}_{11} & {\rm NS}_{21} & {\rm NS}_{21} \\

{\rm NS}_7 & {\rm NS}_5 & {\rm NS}_{16} & {\rm NS}_7 & {\rm NS}_{22} & {\rm NS}_{22} + {\rm NS}_{23} \\

& & {\rm NS}_{17} & {\rm NS}_{15} & {\rm NS}_{22} & {\rm NS}_{19} + {\rm NS}_{20} \\

& & &  & {\rm NS}_{24} & {\rm NS}_{18} \\
& & &  & {\rm NS}_{25} & {\rm NS}_{16} \\
& & &  & {\rm NS}_{26} & {\rm NS}_{12} 
\end{array}
$$

In addition, some of these orbits match even those of $1^6$ theory:
$$
\begin{array}{r|c|l}
1^2 4^2 ~{\rm theory} & 4^3 ~{\rm theory} & 1^6 ~{\rm theory} \\
\hline
{\rm NS}_{1}+ 2~{\rm NS}_{16} & {\rm NS}_1+ 3~{\rm NS}_7 & {\rm NS}_1 \\
{\rm NS}_3= {\rm NS}_4= {\rm NS}_5= {\rm NS}_6 +{\rm NS}_{20} & {\rm NS}_3 + {\rm NS}_{13} & {\rm NS}_2 \\
{\rm NS}_8= {\rm NS}_9= {\rm NS}_{10} & {\rm NS}_{14} =3~{\rm NS}_{10} & 3~{\rm NS}_3 \\
{\rm NS}_{11}+ {\rm NS}_{12} & 3~({\rm NS}_8+ {\rm NS}_9 ) & 3~{\rm NS}_3 \\
{\rm NS}_{13}= {\rm NS}_{14}= {\rm NS}_{15} & {\rm NS}_{15}= {\rm NS}_{12} + 2~{\rm NS}_{11} &  {\rm NS}_4 \\
\end{array}
$$

\appendix
\section{Appendix: Theta functions of given characteristic} \label{sec:theta-functions}

A holomorphic function $T:\C \to \C$ is called a {\em theta
  function with period $\tau$ and characteristic $(a_1,b_1;
  a_2,b_2)$} if it is almost periodic on the lattice, ie if it
  transforms according to
\begin{equation}
  T(v+1)= e^{a_1 v +b_1} T(v), \qquad {\rm and } \qquad
  T(v+\tau)= e^{a_2 v +b_2} T(v)
\end{equation}
We call $n:=(a_1 \tau -a_2)/2\pi i$ the {\em degree} of the
function. 

For example, the following functions are all theta functions with
characteristic and degree
\begin{equation*}
  \begin{array}{rll}
       y^{1/2}:& \quad (0,i\pi;0,i\pi\tau) &0 \\
   \th_1(v):&\quad (0,i\pi;-2\pi i, -i\pi(\tau+1)) &1  \\
   \th_2(v):&\quad (0,i\pi;-2\pi i, -i\pi\tau)  &1\\
   \th_3(v):&\quad (0,0;-2\pi i, -i\pi\tau) &1 \\
   \th_4(v):&\quad (0,0;-2\pi i, -i\pi(\tau+1)) &1 \\
   \th_1(2v|2\tau):&\quad (0,0;-4\pi i, -2\pi i\tau-i\pi) &2 \\
   \th_2(2v|2\tau):&\quad (0,0;-4\pi i, -2\pi i\tau) &2 \\
   \th_3(2v|2\tau):&\quad (0,0;-4\pi i, -2\pi i\tau) &2 \\
   \th_4(2v|2\tau):&\quad (0,0;-4\pi i, -2\pi i\tau-i\pi) &2 \\
   \th_i(v)^2:&\quad (0,0;-4\pi i, -2\pi i\tau) &2,\quad i=1,\dots,4
  \end{array}
\end{equation*}
Note that characteristics add up when multiplying theta functions.
Note also that $\th_3(nv|n\tau)$ and $\th_3(v|{\tau\over n})$ are of
  degree $n$ and characteristic $(0,0;-2n\pi i, -n\pi i\tau)$.  As
  another example, consider the character functions of the level $k$
  and isospin $l$ representation of affine $su(2)$ algebra
  \cite{ET-88-2}:
\begin{equation}
  \chi_k^l(y):= {q^{(l+1/2)^2/(k+1/2)-1/8} \over \prod_{n\geq
  1}(1-q^n)(1-y^2 q^n)(1-y^{-2} q^{n-1}) } \sum_{m\in\Z}
  q^{(k+2)m^2+(2l+1)m} \left( y^{2m(k+2)+2l} - y^{-2m(k+2)-2l-2} \right) 
\end{equation}
This is a theta function of characteristic $(0,0;-4k\pi i,
-2ki\pi\tau)$ and degree $2k$, ie it transforms like\\
$\chi_k^l(v+\tau) = q^{-k} y^{-2k} \chi_k^l(v)$.

Each theta function can be multiplied by trivial theta functions (ie
of degree 0) so that the resulting characteristic reads $(0,0;-2\pi in,
b_2) $ where $n$ the degree (an integer).  For fixed $b_2$, this is a
vector space of dimension $n$ as can be seen from the fact that
contour integration around one lattice cell yields $n$ zeros for $T$:
$P-Z=\oint T'/T =\oint \p \log T =-n$.  We denote this complex
vector space by $\CT_{n,b_2}$.  For $b_2=-n\pi i\tau$, it's spanned by $
\th_3(nv|n\tau), y~\th_3(nv+\tau|n\tau),\dots, y^{n-1}
~\th_3(nv+(n-1)\tau|n\tau)$. 

Thus for instance, all degree 2 theta functions of characteristic
$(0,0;-4\pi i, -2\pi i\tau)$ should be expressible as linear
combinations of $\th_1(v)^2$ and $\th_3(v)^2$ (or any two of the
$\th_i(v)^2$, $i=1,\dots,4$) with $\tau$-dependent coefficients.  This
was the case for the N=4 massless NS characters (\ref{NS-char}), for
$\th_2(v)^2$ or $\th_4(v)^2$ as in (\ref{abstruse-general}), or for the
level 1 $su(2)
$ theta functions:
\begin{equation}
  \begin{split}
     \chi_1^0(y):= {q^{-1/24} \over \prod_{n\geq 1}(1-q^n)(1-y^2
     q^n)(1-y^{-2} q^{n-1})}  \sum_{m\in\Z}
  q^{3m^2+m} \left( y^{6m} -y^{-6m-2} \right) &= {\th_3(2v|2\tau) \over
     \eta}\\
     \chi_1^{1/2}(y):= {q^{-5/24} \over \prod_{n\geq 1}(1-q^n)(1-y^2
     q^n)(1-y^{-2} q^{n-1})}  \sum_{m\in\Z}
  q^{3m^2+m} \left( y^{6m+1} -y^{-6m-3} \right)&= {\th_2(2v|2\tau) \over
     \eta}. \\
  \end{split}
\end{equation}
The right hand sides can be obtained by noting that these too belong
to $\CT_{2,-2\pi i\tau}$ (and by checking the equalities at
$y=1,~q^{1/2}$ say).  Alternatively, they are reproduced by the
quintuple identity (\ref{quintuple}).  

Similarly, any element of $\CT_{2,-2\pi i\tau}$ can be spanned by the
N=4 characters $\hat{\rm ch}_{0,\half}^{\rm NS}$ and ${\rm ch}_0^{\rm
  NS}$, as was done with the NS orbits in (\ref{NS-orbits}).

\section{Appendix: Formulae for theta functions}\label{app-theta}

These are standard definitions and formulae for theta functions. Some
of this material is drawn from Appendix A of \cite{K-97}.

\centerline{\bf Definition}

\be
\th[^a_b](v|\tau)=\sum_{n\in Z}q^{{1\over 2}\left(n-{a\over
2}\right)^2}
e^{2\pi i\left(v-{b\over 2}\right)\left(n-{a\over 2}\right)}
\,,\label{t1}\ee
where $a,b$ are real and $q=e^{2\pi i\tau}$.  We also set
$y=e^{2\pi iv}$.

\vskip .5cm
\centerline{\bf Periodicity properties}

\be
\th[^{a+2}_{\phantom{+}b}](v|\tau)=\th[^a_b](v|\tau)\;\;\;,\;\;\;
\th[^{\phantom{+}a}_{b+2}](v|\tau)=e^{i\pi a}\th[^a_b](v|\tau)
\,,\label{t2}\ee
\be
\th[^{-a}_{-b}](v|\tau)=\th[^{a}_{b}](-v|\tau)\;\;\;,\;\;\;
\th[^a_b](-v|\tau)=
e^{i\pi ab}\th[^a_b](v|\tau) ~~~(a,b\in Z)
\,.\label{t3}\ee

In the usual Jacobi/Erderlyi  notation we have $\th_1=\th[^1_1]$,
$\th_2=
\th[^1_0]$, $\th_3=\th[^0_0]$, $\th_4=\th[^0_1]$.

\vskip .5cm
\centerline{\bf Product formulae}

\begin{equation}  \label{t6}
  \begin{array}{ll}
\th_1(v|\tau)= -i \sum_{n\in\Z} (-1)^n q^{(n+\half)^2/2} y^{n+\half} 
   &= 2q^{1\over 8}\sin[\pi v]\prod_{n=1}^{\infty}
   (1-q^n)(1-q^n y)(1-q^n y^{-1})\\
\th_2(v|\tau)=\sum q^{(n+\half)^2/2} y^{n+\half}
&=2q^{1\over 8}\cos[\pi v]\prod (1-q^n)(1+q^n y)(1+q^n y^{-1})\\
\th_3(v|\tau)= \sum q^{n^2/2} y^n
&=\prod (1-q^n)(1+q^{n-1/2} y)(1+q^{n-1/2} y^{-1})\\
\th_4(v|\tau)= \sum (-1)^n q^{n^2/2} y^n 
&=\prod (1-q^n)(1-q^{n-1/2} y)(1-q^{n-1/2} y^{-1})
  \end{array}
\end{equation}

Define also the Dedekind $\eta$-function:
\be
\begin{split}
  \eta(\tau) &= q^{1\over 24}\prod_{n=1}^{\infty}(1-q^n) 
= q^{1\over 24}~\th_4(\tau/2|3\tau) 
= -i q^{1\over 6}~\th_1(\tau|3\tau) 
= {1\over\sqrt{3}} ~\th_2({1\over 6}|{\tau\over 3})  \\
&= q^{1\over 24} \sum_\Z (-1)^n q^{n(3n+1)/2}
\end{split}
\,.\label{t10}\ee
It is related to the $v$ derivative of $\th_1$:
\be
\left. {\partial\over \partial v} \right|_{v=0}\th_1(v) =: \th_1'=2\pi
~\eta^3(\tau) 
\label{t11}\ee
and satisfies
\be
\eta\left(-{1\over \tau}\right)=~\sqrt{-i\tau}~\eta(\tau)
\,.\label{tt1}\ee
The other $v$-derivatives yield (at $v=0$):
\be
\th_1''=0 = \th_2'=\th_3'=\th_4'
\,.\label{tt2}\ee

\vskip .5cm
\centerline{\bf $v$-periodicity formula}

\begin{equation} \label{t12}
    \th[^a_b]\left(v+{\epsilon_1\over 2}\tau+{\epsilon_2\over 2}
    \bigg|\tau\right) = e^{-{i\pi\tau\over
    4}\epsilon_1^2-{i\pi\epsilon_1 \over 2}(2v-b)- {i\pi \over
    2}\epsilon_1\epsilon_2}~\th[^{a-\epsilon_1}_{b-\epsilon_2}](v|\tau) 
\end{equation}
\begin{equation}
  \begin{array}{rl}
  \th[^a_b](v+\half) &=  \th[^a_{b-1}](v) \\
\th[^a_b](v+{\tau\over 2}) &=  i^b q^{-1/8} y^{-1/2}~\th[^{a-1}_b](v) \\
\th[^a_b](v+{1+\tau\over 2}) &= -i^{b+1} q^{-1/8} y^{-1/2}~\th[^{a-1}_{b-1}](v)\\
  \th[^a_b](v+1) &= (-1)^a ~\th[^a_b](v) \\
\th[^a_b](v+\tau) &=  (-1)^b q^{-1/2} y^{-1}~\th[^a_b](v)
  \end{array}
\end{equation}
That is, $\th[^a_b]$ is a theta function of characteristic $(0,i\pi
a;-2\pi i, -i\pi(\tau+b) )$ and degree 1, see appendix
\ref{sec:theta-functions}.

At the half-periods ($v=0,~\epsilon_{1,2}=0,1$), we are back to the {\em theta
constants} (``Theta Nullwerte''):
\begin{equation}
  \begin{array}{lll}
  \th_1(\half)=\th_2 \qquad&
  \th_1(\frac{\tau}{2})=iq^{-1/8}~\th_4 \qquad &
  \th_1(\frac{1+\tau}{2})= q^{-1/8}~\th_3\\
  \th_2(\half)=0 \qquad&
  \th_2(\frac{\tau}{2})=q^{-1/8}~\th_3 \qquad &
  \th_2(\frac{1+\tau}{2})= -iq^{-1/8}~\th_4\\
  \th_3(\half)=\th_4 \qquad &
  \th_3(\frac{\tau}{2})=q^{-1/8}~\th_2 \qquad &
  \th_3(\frac{1+\tau}{2})= 0\\
  \th_4(\half)=\th_3 \qquad &
  \th_4(\frac{\tau}{2})=0 \qquad &
  \th_4(\frac{1+\tau}{2})= q^{-1/8}~\th_2\\
  \end{array}
\end{equation}
At the quarter periods, we have 
\begin{equation}
  \th_4({\tau\over 4})= -i~\th_1({\tau\over 4}) =
  q^{-\half}~{\eta(\tau/4)~\eta\over\eta(\tau/2)} \qquad \qquad
  \th_2({\tau\over 4})= \th_3({\tau\over 4}) =
  q^{-\half}~{\eta(\tau/2)^2 \over\eta(\tau/4)}
\end{equation}

\vskip .5cm
\centerline{\bf Useful identities}

\begin{equation} \label{theta-eta}
  \th_2 = 2~{\eta(2\tau)^2\over\eta} \qquad  \qquad 
  \th_3 = {\eta^5\over\eta(2\tau)^2~\eta(\tau/2)^2}  \qquad    \qquad 
  \th_4 = {\eta(\tau/2)^2\over\eta}
\end{equation} 
\begin{equation} \label{t13}
  \begin{array}{rl}
\th_2 \th_3 \th_4 &= 2~\eta^3 \\  
\th_3(z|\tau)~\th_3(z'|\tau) + \th_2(z|\tau)~\th_2(z'|\tau)
&= \th_3({z+z'\over 2}|{\tau\over 2})~\th_3({z-z'\over 2}|{\tau\over 2}) \\
\th_3(z|\tau)~\th_3(z'|\tau) - \th_2(z|\tau)~\th_2(z'|\tau)
&= \th_4({z+z'\over 2}|{\tau\over 2})~\th_4({z-z'\over 2}|{\tau\over 2}) \\
\th_2(v|\tau)^4-\th_1(v|\tau)^4 &=\th_3(v|\tau)^4-\th_4(v|\tau)^4
  \end{array}
\end{equation}
For $v=0$, the latter is but Jacobi's {\em abstruse identity}: 
\begin{equation} \label{abstruse}
  \th_3^4 =\th_2^4 +\th_4^4.  
\end{equation}

A more elaborate formula is the {\em quintuple identity}:
\begin{equation}\label{quintuple}
  \begin{split}
    \sum_{n\in\Z} (-1)^n & q^{(3n^2+n)/2} (y^{3n+\half}+y^{-3n-\half}) \\
  &= (y^\half+y^{-\half}) \prod_{n\geq 1}(1-q^n)(1+yq^n)(1+y^{-1}q^{n})
  (1-y^2q^{2n-1})(1-y^{-2}q^{2n-1}) \\
  &=  (y^\half+y^{-\half}) \prod_{n\geq 1} (1-q^n)\frac{(1-y^2
  q^n)(1-y^{-2}q^n)}{(1-yq^n)(1-y^{-1}q^n)}
  \end{split}
\end{equation}
Here are few instances of {\em Riemann addition formulae}:
\begin{equation}\label{riemann-addition}
  \begin{split}
    \th_3(u+v) \th_3(u-v) \th_3^2 &= \th_3(u)^2 \th_3(v)^2 +
    \th_1(u)^2 \th_1(v)^2  \\
    &= \th_4(u)^2 \th_4(v)^2 +\th_2(u)^2 \th_2(v)^2 \\
    \th_4(u+v) \th_4(u-v) \th_4^2 &= \th_3(u)^2 \th_3(v)^2 -
    \th_2(u)^2 \th_2(v)^2  \\
    &=\dots
  \end{split}
\end{equation}
About twenty such formulae can be found on p.20 of \cite{M-82}.  As
special cases ($u=0,1/2$), we recover formulae that generalise Jacobi's
abstruse identity:
\begin{equation}\label{abstruse-general}
  \begin{split}
    \th_3(v)^2 \th_3^2 = \th_2(v)^2 \th_2^2 +\th_4(v)^2 \th_4^2,\\
    \th_4(v)^2 \th_3^2 = \th_3(v)^2 \th_4^2 +\th_1(v)^2 \th_2^2.
  \end{split}
\end{equation}

\vskip .5cm
\centerline{\bf Duplication formulae}

\be \label{dupl-1}
\th_2(0|2\tau)={1\over \sqrt{2}}\sqrt{\th^2_3-\th^2_4}
\;\;\;,\;\;\;
\th_3(0|2\tau)={1\over \sqrt{2}}\sqrt{\th^2_3+\th^2_4}
\,,\ee
\be \label{dupl-2}
\th_4(0|2\tau)=\sqrt{\th_3\th_4}
\;\;\;,\;\;\;
\eta(2\tau)=\sqrt{{\th_2~\eta\over 2}}
\,.\ee
The last two of these are readily seen, while the first two follow from
(\ref{theta-eta}) and from the next properties (most can be derived using the
product form for $\th $):
\begin{equation}
  \begin{split}
    \th_2 &= 2 q^{1/8}  \th_2(\tau|4\tau) = 2 q^{1/8}\th_3(\tau|4\tau)\\
    \th_3(v|\tau) &= \th_3(2v|4\tau) + \th_2(2v|4\tau) \\
    \th_4(v|\tau) &= \th_3(2v|4\tau) - \th_2(2v|4\tau)\\
    \th_3^2-\th_2^2 &= \th_4(0|\tau/2)^2 \\
    \th_3^2+\th_2^2 &= \th_3(0|\tau/2)^2 \\
  \end{split}
\end{equation}
\begin{equation}
  \begin{array}{llll}
\th_2  \th_3 &= \half  \th_2(0|{\tau\over 2})^2 &= 2 q^{1/8}~ \th_3({\tau\over 2}|2\tau)^2 &= 2 \left( \eta^2/\eta({\tau\over 2}) \right)^2 \\
    \th_2  \th_4 &= q^{-1/8}~\th_2({1\over 4}|{\tau\over 2})^2 &= 2 q^{1/8}~
    \th_3({1+\tau\over 2}|2\tau)^2 &= 2\left(\eta({\tau\over 2})~\eta(2\tau)/\eta \right)^2 \\
    \th_3 \th_4 &= \th_3({1\over 4}|{\tau\over 2})^2 &=
    \th_4(0|2\tau)^2 &= \left( \eta^2/\eta(2\tau) \right)^2 
  \end{array}
\end{equation}

\vskip .5cm
\centerline{\bf Heat equation}

The $\th$-functions satisfy the following heat equation
\be
\left[{1\over (2\pi i)^2}{\partial^2\over \partial v^2}-{1\over
i\pi}{
\partial\over \partial\tau}\right]\th[^a_b](v|\tau)=0,
\label{t16}\ee
as well as
\be
{1\over 4\pi i}{\th_2''\over
\th_2}=\partial_{\tau}\log\th_2={i\pi\over 12}
\left(E_2+\th_3^4+\th_4^4\right)
\,,\label{t16-bis}\ee
\be
{1\over 4\pi i}{\th_3''\over
\th_3}=\partial_{\tau}\log\th_3={i\pi\over 12}
\left(E_2+\th_2^4-\th_4^4\right)
\,,\label{t16-biss}\ee
\be
{1\over 4\pi i}{\th_4''\over
\th_4}=\partial_{\tau}\log\th_4={i\pi\over 12}
\left(E_2-\th_2^4-\th_3^4\right)
\,,\ee
where the $E_{2}$ is the second Eisenstein series.  We note that
(\ref{t16-bis}) can be rewritten as 
\be \label{t16-tris}
\partial_{\tau}\log \frac{\th_2}{\eta}= \frac{i\pi}{12}(\th_3^4+\th_4^4)
\,,\ee
and more generally for $(a,b)\neq (1,1)$:
\be
\partial_{\tau}\log \frac{\th[^a_b]}{\eta}=
\frac{i\pi}{12}\left( \th^4[^{a+1}_b]-\th^4[^a_{b+1}] +(-1)^b
\th^4[^{a+1}_{b+1}] \right) 
\,.\ee

\vskip .5cm
\centerline{\bf The Weierstrass function}

\be
\wp(z)=4\pi i\ \p_{\tau}\log~\eta(\tau)-\p^2_z\log\th_1(z)={1\over
z^2}+{\cal O}(z^2)
\label{ww}\ee
is even and is the unique analytic function on the torus with a double
pole at zero.  
\be
\wp(-z)=\wp(z)\;\;\;,\;\;\;\wp(z+1)=\wp(z+\tau)=\wp(z)
\,,\ee
\be
\wp(z,\tau+1)=\wp(z,\tau)\;\;\;,\;\;\;\wp\left({z\over \tau},
-{1\over \tau}\right)=\tau^2~\wp(z,\tau)
\,.\ee
The constant of $z$ in (\ref{ww}) has been chosen so as to cancel the
$z^0$ term in the Laurent expansion, and it equals also
\be
4\pi i\ \p_{\tau}\log~\eta(\tau) = \frac{-\pi^2}{3}\ E_2 = \frac{1}{3} \frac{\th_1'''}{\th_1'}.
\label{ww2}\ee
Alternatively, performing the logarithmic derivative in an appropriate
branch, we can express (\ref{ww}) as
\be \label{wp-theta}
\wp(z)=\left( \frac{\th_1'}{\th_3} \right)^2 \ \left(
\frac{\th_3}{\th_1} (z) \right)^2 + {\rm const}
\ee
where the constant equals $\frac{1}{3} \frac{\th_1'''}{\th_1'}
-\frac{\th_3''}{\th_3} = - 4\pi i \partial_{\tau}\log
\frac{\th_3}{\eta}= {\pi^2 \over 3} (\th^4_2 -\th^4_4)$. 

If we divide the intervals $[0,\tau]$ and $[0,1]$ into $n$
parts and consider the regular grid (on the fundamental lattice)
marked by the points $(s+r\tau)/n$ for $s,r=0,\dots,n-1$, the
$\wp$-values at these points transform into each other under the
action of $SL(2,\Z)$: 
\begin{equation}
  \begin{split}
  \wp\left( {s+r\tau\over n},\tau \right) &\stackrel{T}{\longrightarrow}
  \wp\left( {(s+r)+r\tau\over n},\tau \right)\\
    \wp\left( {s+r\tau\over n},\tau \right) &\stackrel{S}{\longrightarrow}
    \wp\left( {-r+s\tau\over n\tau},-{1\over\tau} \right) = \tau^2
  \wp\left( {-r+s\tau\over n},\tau \right) 
  \end{split}
\end{equation}
Putting all these $\wp$-values -- except $s=r=0$ -- into a vector
with $n^2-1$ components, we obtain a vector-valued modular form of
weight 2.  Summing all components yields 0, as there is no modular
form of weight 2:
\begin{equation}
  \label{wp-grid}
  {\sum_{r,s}}'  \wp\left( {s+r\tau\over n} \right) =0,
\end{equation}
where the prime indicates exclusion of $s=r=0$.  For $n=2$, this is
the well-known identity for the half-periods,
\begin{equation}
  \wp(1/2)+\wp(\tau/2)+\wp((1+\tau)/2) = 0,
\end{equation}
also derivable from (\ref{wp-theta}).

\end{document}